\documentstyle[12pt]{article}
\newcommand{\tiN}{\raisebox{-6.5pt}{$
\displaystyle \stackrel{\displaystyle N}{\sim} $}}
\newcommand{\tiM}{\raisebox{-6.5pt}{$
\displaystyle \stackrel{\displaystyle M}{\sim} $}}
\newcommand{\beq}{\begin{equation}} \newcommand{\eeq}{\end{equation}}
   \makeatletter
   \@addtoreset{equation}{section}
   \makeatother
   
\newtheorem{theorem}{Theorem}[section]
\newtheorem{definition}{Definition}[section]

\begin{document} \title{Canonical Quantization \\  of Spherically Symmetric
Gravity \\ in Ashtekar's Self-Dual Representation} \author{T.\
Thiemann\thanks{Supported by the Graduierten-Programm of the DFG;  E-mail:
thiemann@phys.psu.edu}$~$ and H.A.\ Kastrup\thanks{E-mail: KW010KA@DACTH11} \\
Institut f\"ur Theoretische Physik, RWTH Aachen \\ D-52074 Aachen,
Germany}\date{  }     \maketitle
  \begin{abstract} We show that the quantization of  spherically symmetric
pure gravity can be carried out completely in the framework of Ashtekar's
self-dual representation. Consistent operator orderings can be given for the
constraint functionals yielding two kinds of solutions for the constraint
equations, corresponding  classically to globally nondegenerate or
degenerate
metrics. The physical state functionals can be determined by quadratures and
the reduced Hamiltonian system possesses 2 degrees of freedom, one  of them
corresponding to the classical Schwarzschild mass squared and the
canonically conjugate one representing  a measure for the deviation  of the
nonstatic field configurations from  the static Schwarzschild one. There is
a natural choice for the  scalar product making the 2 fundamental
observables self-adjoint. Finally, a unitary transformation is performed  in
order to calculate the triad-representation of the physical state
functionals and to provide for a  solution of the appropriately
regularized Whee\-ler-DeWitt equation. \end{abstract} \section{Introduction}
The introduction of  new canonical variables for the Hamiltonian framework
of general relativity by Ashtekar \cite{1} has considerably enhanced the
chances of finding a consistent quantum theory for gravity because the new
constraint functionals depend only polynomially on the field variables, in
constrast to the much more complicated ADM-approach. Up to now the effort to
find solutions of the quantum constraint equations has been concentrated
mainly on the loop representation introduced by Rovelli and Smolin. We refer
to several recent review articles \cite{2}-\cite{5} for the  discussion of
the underlying ideas and of the status of solved and unsolved problems. \\
In the following we shall show that Ashtekar's nonperturbative quantization
programme can be carried out completely for spherically symmetric field
configurations by using the  self-dual representation originally suggested
by Ashtekar. \\ The  basic canonical variables in Ashtekar's approach to
quantum gravity are the connection coefficients $A^{i}_a(x) $ as
configuration variables and the densitized triads $ \tilde{E}^{a}_i(x),
a=1,2,3, i=1,2,3$ as momentum variables characterized as follows: \\ Let
$q_{ab}, a,b=1,2,3$ be the spatial metric on the 3-dimensional Cauchy
hypersurface $ \Sigma$, then the cotriad coefficients $E^{i}_a(x) $ obey the
relations \beq q_{ab}(x)=  E^{i}_a(x)\, E^{j}_b(x)\,\delta_{ij},~~~
E^{a}_i(x)\, E^{j}_a(x) = \delta^{j}_i.  \eeq The indices $a,b$ are space
indices and the indices $i,j$ are internal indices with respect to the
Lie-algebra $ {\cal L} (SO(3)): E_a(x) = E^{i}_a(x) \lambda_i, \lambda_i \in
{\cal L}(SO(3)), tr(\lambda_i \lambda_j)= \delta_{ij} $ .\\ The densitized
triad coefficients  $ \tilde{E}^{a}_i(x)$ are defined by \beq
\tilde{E}^{a}_i(x) = \sqrt{q(x)} E^{a}_i(x) ,~~~ q(x) =
\mbox{det}(q_{ab}(x)). \eeq The complex coefficients $ A^{i}_a(x)$ of the
Ashtekar connection $A_a=A_a^{i}\lambda_i$ are the pull-backs to the spatial
surface $\Sigma$ of the self-dual part of the spacetime spin-connection. They
can also  be defined by the linear combination \beq A^{j}_a(x)
\lambda_j=\Gamma^{j}_a(x) \lambda_j + i K_{a}^j(x)
\lambda_j,~~~~ K_{a}^i = K_{ab}E^{b}_j \delta^{ji} ,\eeq where
$\Gamma^{i}_a(x)$ are the
coefficients of the spatial spin connection expressed by the triads and
$K_{ab}$ are
the coefficients of the extrinsic curvature of $\Sigma$ which play the role
of canonical
momenta in the ADM formalism. The last equation may be viewed as a complex
canonical
transformation. \\ The  action for gravity formulated in terms of Ashtekar's
variables takes the form \begin{eqnarray}S & =
&\frac{1}{\kappa}\int_{R^1}\{\Theta_L-(i\Lambda^i\circ{\cal G}_i
-i N^a\circ
H_a+\frac{1}{2}\tiN \circ H)dt\}, \\\Theta_L & = & -i\tilde{E}^a_i\circ
dA_a^i, \\{\cal
G}_i & = & {\cal D}_a \tilde{E}^a_i, \\H_a & = &
\epsilon_{abc}B^c_i\tilde{E}^b_i,\\H & =
& \epsilon_{abc}\epsilon^{ijk} B^a_i \tilde{E}^b_j\tilde{E}^c_k \;.
\end{eqnarray}The
notation is the usual one : $\Lambda^i,N^a$ and $ \tiN$ are the
Lagrange-multipliers
corresponding to the Gauss, vector and scalar constraints ${\cal G}_i ,
H_a $ and $  H$
respectively, that is $\Lambda^i=A_0^i$, whereas
 $N^a$ and  $N=\det(q)^{1/2} \tiN$ are the
usual shift and  lapse functions . The covariant differential with
respect to $A^i$ is
denoted by $\cal D$, $\Theta_L$ is the Liouville-form and the
'magnetic fields' with
respect to the curvature 2-form $F^i$ of the connection $A^i$ are given by
\begin{equation}B^a_i:=\frac{1}{2}\epsilon^{abc}F_{bc}^i.\end{equation}The
convention
\beq f\circ g:=\int_\Sigma d^3x f(x)g(x)  \eeq has been used, where f(x)g(x)
 is a
scalar density of weight 1.\\ The paper is organized as follows: Section
2 contains the
basic formulas for the theory concerning spherically symmetric field
configurations only. \\
Due to their  $SO(3)$ transformation properties the
connection variables and the triads can
be expressed by 6 functions $A_I(t,x),  $ and $E^I(t,x), I=1,2,3$.   Here
$t$ is a "time"
variable and x is a (local) spatial variable which becomes the usual
 Euclidean
radial variable $r$ at spatial infinity. \\ The metric $(q_{ab})$ now takes
the form \beq
(q_{ab}) =\mbox{diag}(\frac{E}{2E^1}, E^1, E^1 \sin^2\theta ),~~~~
\det(q_{ab})=
\frac{1}{2} E^1 E \sin^2\theta,~~~ E= (E^2)^2 + (E^3)^2 , \eeq which shows
that
the variables $E^1$ and $E$  determine the sign and the degeneracies of the
spatial metric.
\\The 4-dimensional line element is  given by \beq ds^2 = -(N(x,t)\,dt)^2 +
q_{xx}(x,t)(dx+N^x(x,t) \,dt)^2 +q_{\theta \theta}(t,x) (d\theta^2 + \sin^2
\theta d\phi^2)\; .  \eeq Section 3 discusses properties of the classical
phase space,
especially the fall-off properties of different quantities for asymptotically
flat
spaces motivating the choice of function spaces on which the quantum states
have support.
\\Section 4 is the central part of the paper giving  the solutions of the
constraint
equations in the self dual connection (Schr\"odinger) representation,
where \beq
\hat{A}_I(x)\Psi[A_I] = A_I (x) \Psi[A_I],~~~~\hat{E}^I \Psi[A_I]  =
\frac{\delta}{\delta
A_I} \Psi[A_I],\eeq $\Psi[A_I] $ being a holomorphic functional of the
classical field
variables $A_I$ on which the multiplication operators $ \hat{A}_I  (x)$
and the functional
derivative operators $\hat{E^I} $ act in the  way just described. The crucial
point as to
the solution of the constraint equations is that they can be transformed
into equivalent
ones containing the functional derivatives with respect to $A_I$ at most
linearly. \\ Two
types of solutions emerge:\\ The first one, \beq\Psi_I [A_J] =\Psi(Q) ,
{}~~~~ Q =-i
\int_{\Sigma}dx \frac{A_2B^2+A_3B^3}{A} (B^1)^{-2} \;.  \eeq corresponds
to configurations
which classically have a singular metric at most at isolated points. Here
\begin{eqnarray}
B^1 & = & (A-2)/2, \\ & &  A=(A_2)^2+(A_3)^2, \nonumber  \\  B^2 & = &
A^{\prime}_3+ A_1
A_2, ~~~ f^{\prime}=\frac{df(x)}{dx}, \\  B^3 & = & -A^{\prime}_2 + A_1A_3,
\end{eqnarray}are the "magnetic fields" associated with the spherically
reduced connection represented by the coefficients $A_I$ and $\Psi(Q)$ is
any smooth
function of Q. \\ The integrand in the quantity $Q$  bears a strong
resemblance to the
integrand in the Chern-Simons functional \beq\int_{\Sigma}dx
\frac{A_2B^2+A_3 B^3}{A}B^1
\eeq   of the theory. \\ The second type of solutions, \beq
\Psi_{II}[A_J]=\Psi(\int_{\Sigma} dx A_1 (x)),\eeq belongs to
configurations which  have
 globally singular metrics, namely E=0.
 \\ We shall mainly be concerned with the first type
of solutions: \\ The gauge and diffeomorphism invariant quantity $Q$ is
weakly real and
emerges as one of the 2 basic observables of the system.
\\  The second, canonically conjugate
observable P is \beq P= \int_{\Sigma} dxT(x)(B^1(x))^2 E^1(x),
\int_{\Sigma}dx T(x)=1,
\eeq - $T(x)$ is a test function -  which acts as a
derivative operator $ \hat{P}=-i d/dQ $
on the state functionals $\Psi(Q)$
on which $\hat{Q}$ acts as multiplication operator, that
is we have  \beq[\hat{Q},\hat{P}]=i. \eeq   The Hilbert space of physical
states consists
of all square-integrable functionals $\Psi(Q)$ with the scalar product
\beq<\Psi_1,\Psi_2>
= \int_{R} dQ \bar{\Psi}_1(Q) \Psi_2(Q). \eeq This form of the scalar
product is almost
obvious but can be justified in detail by BRST-type arguments.
%\\ The ADM-energy-operator
%$\hat{e}_{ADM}$ acts on the states $\Psi(Q)$ as \beq \hat{e}_{ADM}\Psi(Q)
= const.\
%\frac{d^2 \Psi(Q)}{dQ^2},\eeq where the constant  depends on the
configuration.
\\ Section 5 discusses the problem why there are 2 observables, P and
Q because, according
to Birkhoff's theorem concerning the uniqueness of the Schwarzschild
solutions, one
expects
just 1 physical degree of freedom for spherically symmetric gravity,
namely the
Schwarzschild mass $m$. The reason for the occurence of a second observable,
$Q$, is
related to that discussed in reference \cite{6}: \\ For {\em stationary}
classical spacetimes
the quantity $Q$ can be expressed as \beq Q = \frac{1}{4c}\int_{\Sigma} dx
\frac{N^x}{\tiN}\frac{(E^1)^{\prime}}{E^1(1+\sqrt{\frac{E^1}{c})}} \; .
\eeq The real
parameter $c$ is a constant of integration which takes the
value $16 m^2, m:$ Schwarzschild
mass, for the Schwarzschild solution. For the standard Schwarzschild
solution associated with a {\em static} foliation (slicing of space
and time)
 one gets
$Q=0$ because $N^x=0$ here  and $P = 4m^2$. If $Q \neq 0$ then we must
have $N^x \neq 0$,
i.e. we cannot have purely static field configurations. Here the
following difference
between
the notions of gauge in the Lagrangean and the Hamiltonian formulation
respectively
comes into play \cite{6}: \\ While one uses any spacetime-diffeomorphism to
relate gauge-equivalent metrics in the Lagrangean formulation, only
those diffeomorphisms that are generated by the constraint functionals
define
gauge transformations in the Hamiltonian formulation. Thus, according to
Birkhoff one can
make $N^x$ vanish by an
appropriate spacetime gauge, but this is in general not possible in
the
Hamiltonian framework.\\ Section 6 presents the solution of the constraints
in the triad
representation where the operators $\hat{A}_I$ act as functional derivative
operators and
the operators $\hat{E}^I$ as multiplication operators  on the functionals
$\tilde{\Psi}[E^I]$. The physical functionals $\tilde{\Psi}_c[E^I]$ can
be obtained
explicitly either by transforming the constraints in the triad representation
into a
linearized form (now with respect to  $A_I$) or by performing a (unitary)
Laplace
transformation of the physical connection functionals $\Psi_c(Q)$ into the
triad representation. Both methods yield the same $\tilde{\Psi}_c[E^I]$
which turns out to be a
spherically symmetric solution of the Wheeler-DeWitt equation if the operator
regularization of the latter is appropriately chosen.\section{Definition of
the model}The
reduction of the various classical geometrical
quantities of the Ashtekar formalism  to the
spherically symmetric case has been done in refs. \cite{7}. We here give
mainly the basic
ingredients necessary
for our purposes: \\Let the 3 rotational Killing-fields be denoted by
$L_i$ and the generators of O(3) by $T_i$. Then the reduction is obtained by
imposing\begin{equation} {\cal L}_{L_j} E_i=-[T_j,E]_i \; ,\end{equation}
where on the rhs
the bracket means the commutator in the Lie-algebra of $SO(3)$.
$ {\cal L}_{L_j}$ is the
Lie derivative with respect to $L_j$. Hence it is required that the
rotation of the triads
in the tangent bundle is compensated by an internal rotation in the
$SO(3)$-bundle. Evaluation of eqn. (2.1) yields the general form of the
reduced triads. It
is not difficult  to compute the reduced form of ${\cal L}SO(3)$-valued
vector densities
of weight one and of  covectors as, for example, the densitized triads
$\tilde{E}^a_i$
and the Ashtekar-connection $A_a^i$ respectively, which make up the
basic variables of
the theory : \beq(\tilde{E}^x,\tilde{E}^\theta,\tilde{E}^\phi) =
 (E^1n_x\sin(\theta),\sqrt{\frac{1}{2}}(E^2
n_\theta+E^3n_\phi)\sin(\theta),\sqrt{\frac{1}{2}}(E^2 n_\phi-E^3 n_\theta)),
\eeq\newpage
 \begin{eqnarray} (A_x,A_\theta,A_\phi) & = & (A_1n_x,\sqrt{\frac{1}{2}}(A_2
n_\theta+(A_3-\sqrt{2}) n_\phi), \\  & & ~~~~~~~~~~~~
\sqrt{\frac{1}{2}}(A_2 n_\phi-(A_3-\sqrt{2})
n_\theta)\sin(\theta)) \; .\nonumber\end{eqnarray}Here we have simultanously
introduced
our notation:\\We denote by $\theta,\phi$ the (global) Killing-parameters,
$t$  and $x$
are
the local time and space coordinates. The up to now arbitrary complex
functions
$E^I=E^I(t,x),A_I=A_I(t,x); I=1,2,3$ depend on t and x only. The vectors
$n_x,n_\theta$ and $n_\phi$ are the usual unit vectors on the sphere
but  are to be understood here
as the generators of $SO(3)$ (thereby making use of the
fact that the adjoint- and defining
representation of $SO(3)$ are equivalent), i.e. the basis of ${\cal L}SO(3)$
here is
angle-dependent.\\One now simply inserts this into Ashtekar's action of
full gravity (1.4),
integrates out the angles (in particular the factor $\sin\theta$ contained in
$\tiN$ drops
out) and obtains the basic quantities of the model summarized in the
following list
($\Lambda=\Lambda^i (n_x)^i,{\cal G}={\cal G}_i (n_x)^i; \kappa/8\pi$ is
Newtons constant)
:\\\\Action:
\begin{equation}S=\frac{4\pi}{\kappa}\int_{R^1}\{\Theta_L-[b+\Phi]dt\},
\end{equation}
Liouville-form:
\begin{equation} \Theta_L=-i E^I\circ dA_I,\end{equation}Constraint
functional:
\begin{equation}\Phi=i\Lambda\circ{\cal G}-iN^x\circ H_x+\tiN\circ H,
\end{equation} where \\Gauss
constraint function (i.e.\ the integrand of the functional):
\begin{equation} {\cal G}={\cal D}_x E^1=(E^1)'+
A_2 E^3-A_3
E^2,\end{equation} Vector constraint function:\begin{equation}
H_x=B^2 E^3-B^3 E^2,
\end{equation}Diffeomorphism constraint function:\begin{equation}
\xi=H_x-A_1{\cal
G}=-A_1(E^1)'+(A_2)'E^2+(A_3)'E^3,\end{equation}Scalar constraint
function:\begin{equation}
H=\frac{1}{2}(E^2(2 B^2 E^1+B^1 E^2)+E^3(2 B^3 E^1+B^1E^3)),
\end{equation}Boundary term:
\begin{equation} b=\int_{\partial\Sigma}(q+p+e), \end{equation}O(2)-charge
density:\begin{equation} q=-i\Lambda E^1, \end{equation} ADM-momentum
density:\begin{equation} p=i N^x(A_2 E^2+(A_3-\sqrt{2})E^3), \end{equation}
ADM-energy
density:\begin{equation} e_{ADM}=\tiN (A_2
E^3-(A_3-\sqrt{2})E^2)E^1,\end{equation}Poisson-brackets:\begin{equation}
\{A_I(x),E^J(y)\}=i\delta_{I}^{J}\delta(x,y)\;,\;\{A_I(x),A_J(y)\}=\{E^I(x),
E^J(y)\}=0
\end{equation}Eqs. of (gauge transformations and) motion :\begin{equation}
\frac{d}{d\tau
}A_I=\{A_I,\Phi+b\},\;\frac{d}{d\tau}E^I=\{E^I,\Phi+b\}\; , \end{equation}
explicitely:\begin{eqnarray}\frac{d}{d\tau}A_1 & = & i[(-i\Lambda'+\tiN(B^2
E^2+B^3 E^3)],
 \\\frac{d}{d\tau}A_2 & = & i[(-i\Lambda A_3+i N^x B^3+\tiN(B^2 E^1+B^1
 E^2)],\nonumber
\\\frac{d}{d\tau}A_3 & = & i[(+i\Lambda A_2-i N^x B^2+\tiN(B^3 E^1+B^1
E^3)],\nonumber
\\\frac{d}{d\tau}E^1 & = & -i[-i N^x(A_2 E^3-A_3 E^2)+\tiN (A_2 E^2+A_3
E^3)E^1],
\nonumber \\\frac{d}{d\tau}E^2 & = & -i[+i(\Lambda-N^x A_1)E^3+i(N^x
E^2)'+\tiN(A_1
E^1E^2+\frac{1}{2}E A_2) \nonumber \\&  & +(\tiN E^1 E^3)'], \nonumber
\\\frac{d}{d\tau}E^3 & = & -i[-i(\Lambda-N^x A_1)E^2+i(N^x E^3)'+\tiN(A_1
E^1E^3+\frac{1}{2}E A_3) \nonumber \\&  & -(\tiN E^1 E^2)']\; .\nonumber
\end{eqnarray}The
classical canonical constraint algebra is:\begin{eqnarray}\{\Lambda_1
\circ{\cal
G},\Lambda_2\circ{\cal G}\}&= & 0,  \\\{N\ast\xi,\Lambda\circ{\cal G}\}
& = &-i
N\Lambda'\circ\cal G, \nonumber\\ \{\tiN \circ H,\Lambda\circ{\cal G}\}
& = &0, \nonumber
\\\{M\circ\xi,N\circ\xi\} & =& i(M N'-M'N)\circ\xi, \nonumber \\
\{M\circ\xi,\tiN\circ H\} & = & i(M\tiN'-M'\tiN)\circ H, \nonumber \\
\{\tiM\circ H,\tiN\circ H\} & = &
i(\tiM\tiN'-\tiM'\tiN)\circ(E^1)^2 H_x \;. \nonumber\end{eqnarray}\\For the
model the
"magnetic fields" take the form\begin{eqnarray}(B^x,B^\theta,B^\phi) & = &
(B^1n_x\sin(\theta),\sqrt{\frac{1}{2}}(B^2 n_\theta+B^3
n_\phi)\sin(\theta),\sqrt{\frac{1}{2}}(B^2 n_\phi-B^3 n_\theta)),\nonumber
 \\(B^1,B^2,B^3) & = &(\frac{1}{2}((A_2)^2+(A_3)^2-2),A_3'+A_1 A_2,-A_2'
 +A_1A_3)  \; ,
 \end{eqnarray}A prime denotes derivation with respect to x, and $\circ$ -
compare eq.\ (1.10) - denotes now an integral over $\Sigma^1$ with local
variable $x$
only,
i.e.\ the integral over the unit sphere has been carried out.\\In order
to make contact
with ref. \cite{7}, one has
to exchange the labels I=2 and I=3 and to replace $A_3+\sqrt{2}$
there by $A_3$ in order  to get $A_3$ here.
 {\em This shift  of $A_3$ by $\sqrt{2}$ on the other
hand is essential to read off the model's kinematical part of the gauge
group from the
table as $O(2)\times Diff(\Sigma^1)$}, which seems to have been overlooked
in ref.
\cite{7}. \\ It is then obvious from the above list that there is a strong
similarity
between our model and full gravity, so that one can hope to learn
something from the model,
valid for full gravity : the diffeomorphisms have been  frozen to the
x-direction, the
internal rotations to the $n_x$-direction. $A_1$ plays the role of an
$O(2)$ gauge
potential, $E^1$ is $O(2)$-invariant while the vectors $(E^2,E^3),(A_2,A_3)$
transform
according to the defining representation of $O(2)$. $A_1,E^2,E^3$ are
densities of weight
one, while $E^1,A_2,A_3$ are scalars. This can be seen from the
transformation formulas
(2.17).\\Moreover, the constraint algebra is again first  class and, as
one can show,  of rank
1 in BRST-terminology, too
(see ref.\ \cite{8}).\\ Remark: Although the shift by $\sqrt{2}$ drops
out from derivatives of $A_3$, the fall-off conditions discussed in ref.
\cite{7}
are now to be imposed on
$A_3-\sqrt{2}$ (see next section). \\Frequent use will be made of the
following
abbreviations :\begin{eqnarray}& &
\alpha:=\arctan(\frac{A_3}{A_2}),~~~\eta:=\arctan(\frac{E^3}{E^2}),~~~
\beta:=\arctan(\frac{
B^3}{B^2}), \nonumber \\ & &A:=(A_2)^2+(A_3)^2,~~~E:=(E^2)^2+(E^3)^2,~~~
B:=(B^2)^2+(B^3)^2
\; .\end{eqnarray}The metric is given by\begin{equation}(q_{ab})=
\mbox{diag}(\frac{E}{2
E^1},E^1,E^1\sin^2(\theta)) \;.\end{equation}For the discussion of the
reality  conditions
the reduction to spherical symmetry of the spin connection is
needed:\begin{eqnarray}(\Gamma_x,\Gamma_\theta,\Gamma_\phi) & = &(\Gamma_1
n_x,\sqrt{\frac{1}{2}}(\Gamma_2 n_\theta+(\Gamma_3-\sqrt{2})n_\phi),
\\ & & ~~~~~~~~~~~~
\sqrt{\frac{1}{2}}(\Gamma_2 n_\phi-(\Gamma_3-\sqrt{2})n_\theta)\sin(\theta)),
\nonumber
\\(\Gamma_1,\Gamma_2,\Gamma_3) & = & (-\eta',-(E^1)'\frac{E^3}{E},(E^1)'
\frac{E^2}{E}) \;
,~~~\eta'= \frac{E^2 E^{3 \prime} -E^3 E^{2\prime}}{E} \;, \end{eqnarray}
and for later use
in section 5 we record the spherically symmetric
reduction of the extrinsic curvature (which
is a Lie-algebra-valued covector like $A_a^i$):\begin{eqnarray}
(K_1,K_2,K_3)& =&
\frac{1}{\tiN E E^1}(E^1(\dot{q}_{xx}-(q_{xx})'N^x-2q_{xx}(N^x)', \\ &
 &~~~~~~~~~~~~~~~ E^2(\dot{E}^1-N^x(E^1)'),
E^3(\dot{E}^1-N^x(E^1)'))\nonumber \;,\end{eqnarray} where
the dot means the derivative with respect to the variable
$t$.

\section{Definition of the phase space}

The following analysis is a simplification of that given in reference
\cite{7}.
The classical Poisson-brackets can be read off from the Liouville form
(we work with geometrical units and set $\hbar=1$; the factor $4\pi$
from angle-integration is also dropped because it is a common pre-factor of
all spherically symmetric terms ) :
\begin{eqnarray} \{A^I(x),A^J(y)\}=\{E^I(x),E^J(y)\}=0, \\
\{A_I(x),E^J(y)\}=i \delta^J_I\delta(x,y) \;\mbox{for all x,y in}\;
\Sigma \;.\nonumber\end{eqnarray}
We next wish to determine the function spaces to which our basic variables
$A_I$ and $E^I$ belong for the case of asymptotically flat field
configurations of the metric.
We follow here the definition of asymptotic flatness given, for example,
in ref. \cite{9}, that is, in simplified form :
\begin{definition}
i) A Riemannian manifold $(\Sigma^3,q)$ is called asymptotically Euclidean
iff
there exists a compact subset $K^3$ of $\Sigma^3$ so that $\Sigma^3-K^3=
\cup_A \Sigma_A^3$, where any
'end' $\Sigma_A^3$ is diffeomorphic to the complement of a ball B in $R^3$
and
q tends to the Euclidean metric at infinity of $\Sigma_A$ in a way to be
specified.\\
ii) A spacetime $(M=R\times\Sigma,g)$ is called asymptotically flat, if i)
holds
and moreover lapse N and shift $\vec{N}$ tend to their Minkowskian values
in a way to be specified.
\end{definition}
Note that $R^3$ is homeomorphic to $R^+\times S^2$ and $R^3-B$ to $R^{>\rho}
\times S^2,\rho>0$,
where $R^+=\{x\in R;\; x\ge 0\},\;R^{>\rho}=\{x\in R;\;\rho<x<\infty\}$, so
that one encounters the following difficulty
after integrating the action over $S^2$ : \\
According to our definition of asymptotic flatness we have in the spherically
symmetric case
\begin{eqnarray}
\Sigma^3-K^3=(\Sigma^1-K^1)\times S^2 & \cong & \cup_A(R^3-B_A)=(\cup_A\;
R^{>\rho_A})\times S^2. \\
\mbox{Thus  } \Sigma^1-K^1 & \cong & \cup_A R^{>\rho_A} \; ,
\end{eqnarray}where $K^1$ is again a compact subset of $\Sigma^1$, which may
be empty. Thus,if one chooses a spatial topology with only one end
(asymptotic region)
one has  at a Cauchy hypersurface possessing a boundary that does not
 correspond to infinity. In that case one would also have to impose
 boundary conditions at the
finite boundary. We assume that for this situation a consistent choice
of the boundary conditions exists. This problem is, of course, an
artefact of the reduction
process and is of no physical relevance from a 3-dimensional viewpoint.
A possible choice, for example, is that the fields have  support outside
the origin or that
one restricts oneself to (connected) manifolds $\Sigma^1$ such that
$\Sigma^1-K^1$ is diffeomorphic to the union of at least 2 copies of
$R^{>\rho}$.
Thinking of Kruskal's extension of the Schwarzschild solution, the latter
option is physically reasonable. \\In the following, we will drop the
superscript on $\Sigma:=\Sigma^1$ and need only treat the fall-off
at infinity in more detail. \\ \\The usually imposed fall-off conditions
in an asymptotically
cartesian frame are as follows:\begin{equation}
(q_{ab})\to (\delta_{ab}+\frac{1}{r}f_{ab}(\vec{n},t)+\frac{1}{r^{1+
\epsilon}}g_{ab}(\vec{n},t)) \; ,
\end{equation} where here and in what follows the asymptotically
radial coordinate which belongs to the given end has been called r; f
and g are smooth tensors of
the angles and the time coordinate t. \\Translating this into the
model yields ($\epsilon>0$)\begin{equation}
(E^1,E)=(r^2(1+f^1(t)/r+O(r^{-1-\epsilon})),2r^2(1+f(t)/r+O(r^{-1-
\epsilon})))\; , \end{equation} which implies \begin{equation}
(E^2,E^3)=r\sqrt{2}(e^2+f^2(t)/r+O(r^{-1-\epsilon}),e^3+f^3(t)/r+O(r^{-1
-\epsilon}))\; , \end{equation} where $f^I,f$ are the obvious
analogues of the tensors given
in (3.4) for the spherically symmetric case and $e^2,e^3$ are at this
stage complex numbers, subject to the constraint $(e^2)^2+(e^3)^2=1$.
The angle-dependence has
completely dropped out while integrating the action over $S^2$.
\\The fall-off-conditions for the conjugate variables $A_I$ are prescribed
by requiring that the
Liouville-form is well defined
:\begin{equation}(A_1,A_2,A_3-\sqrt{2})=(\frac{1}{r^2}(a_1+\frac{c_1}{
r^\epsilon})+\frac{b_1(t)}{r^{3+\epsilon}},\frac{1}{r}(a_2+\frac{c_2}{
r^\epsilon})+\frac{b_2(t)
}{r^{2+\epsilon}},\frac{1}{r}(a_3+\frac{c_3}{r^\epsilon})+\frac{b_3(t)}{
r^{2+\epsilon}}) \; ,\end{equation} where $\delta a_I,\delta c_I=0$
is a possible choice for
the variations of the leading order coefficients of the connections
(the variations are infinitesimal dynamical or gauge transformations).
Note that it makes a
difference whether one requires $E^I\circ\delta A_I\; \mbox{or}\;
A_I\circ\delta E^I$ to be a finite 1-form on the phase space (this is,
of course, also true for
fullgravity).\\The Lagrange-multipliers are test functions that regularize
the constraint distributions.The induced fall-off properties of the latter
are as follows
:\begin{eqnarray}{\cal G} & \to & 2r+\sqrt{2}(a_2 e^3-a_3 e^2-r\sqrt{2}e^2)
\to 2r(1-e^2)+O(1)\; , \nonumber\\H_x & \to & r^{-1}\sqrt{2}((\sqrt{2}a_1+
a_2)e^2+a_3
e^3)+O(r^{-2})\; , \nonumber\\H & \to & r\sqrt{2}((1-e^2)a_3+e^3(a_2+a_1
\sqrt{2}))+O(1) \; .\end{eqnarray} If one insists on a non-vanishing
ADM-energy for improper
gauge transformations (the terminology is explained, for instance, in
\cite{7}; roughly speaking, proper gaugetransformations act as identity on
the constraint
surface of the phase space which implies that in this case the boundary
term $b$ in eqn. (2.11) vanishes) inspecting the boundary term b shows
that one has to choose
the following fall-off propertie sof the test functions that appear in
the action : \\Proper gauge transformations :
\begin{equation}(\Lambda,N^x,\tiN)=(O(r^{-(2+\epsilon)}),O(r^{-\epsilon}),
O(r^{-(2+\epsilon)}),\end{equation} Improper gauge transformations :
\begin{equation}(\Lambda,N^x,\tiN)=(O(r^{-2})\; \mbox{or}\; O(r^{-(2+
\epsilon)}),O(1),O(r^{-2}) .\end{equation} This choice simultanously makes
the action functional
finite provided one imposes the following additional conditions :
\begin{equation} a_1\sqrt{2}+a_2=0,\; a_3(1-e^2)=0 \; , \end{equation}
where the stronger fall-off of
$\Lambda$ corresponds to the choice $a_3=0$. Note that one usually
(see ref. \cite{3}) does not allow the O(3)-charge in full gravity to
be non-vanishing, but it
turns out that for the Schwarzschild solution only the weaker fall-off
is possible because $a_3$ is essentially the Schwarzschild mass,
so that its analogue,
the O(2)-charge, does not vanish, while $a_1,a_2$ turn out to vanish
seperately (see ref. \cite{7}).Formula (3.11) then suggests to restrict
further
:\begin{equation} a_1=a_2=0,\sqrt{2}c_1+c_2=0 \; , \end{equation}which
 will prove essential for the existence of physical states.

\section{Realization of the Quantization Programme}

\subsection{Step 1 : Definition of the $\star-\mbox{algebra}\;\cal A$}

We will now carry out all the steps of the quantization programme as
proposed
by Ashtekar and discussed at length in refs. \cite{2}-\cite{5}.\\
\\
In the first step one is asked to state the equal time canonical
commutation relations (CCR)
as well as the reality conditions ($\star$-relations) which defines an
abstract
$\star$-algebra. The algebra $\cal A$ we choose is the natural one
suggested by the Liouville-form
and is defined by
\begin{eqnarray}
{[}\hat{A}_I(x),\hat{E}^J(y)] & = &-\delta^J_I\delta(x,y), \\
{[}\hat{A}_I(x),\hat{A}_J(y)] & = & [\hat{E}^I(x),\hat{E}^J(y)]=0, \\
(E^I(x))^\star-E^I(x) & = & 0, \\
(A_I(x)-\Gamma_I(x))^\star+(A_I(x)-\Gamma_I(x)) & = & 0 \; .
\end{eqnarray}
However, we will not use the just defined $\star$-relations (4.3) and (4.4)
in order to find the scalar-product (or induced ones on other operators,
even not
in their polynomial version, see, for instance, ref. \cite{10}) for the
following reason : \\
Normally (\cite{2}-\cite{5}) one argues that the inner product is to be
determined by imposing
the above $\star$-relations to become adjointness-conditions with respect
 to
the scalar product. The integration measure of this scalar product will
then in general be highly non-trivial in order to account for the
complicated adjointness
conditions (4.3) and (4.4).
However, even if one succeeds to construct such a measure  this  will in
general not  imply that
the induced adjointness-relations on {\em quantum-observables}
reflect the reality-conditions of their classical counterparts,
i.e. even if an observable is classically real and therefore satisfies a
 necessary
condition to be measurable it is far from granted that the associated
quantum
operator becomes self-adjoint (example: 1-dimensional quantum mechanics;
$O=qp$
is classically real but in quantum theory it fails to be self-adjoint).
Furthermore, there will in general exist non-polynomial observables (as for
example
the quantity Q in this model). So, imposing the adjointness-conditions (4.3)
and (4.4) will then
result in mathematically horrible objects. It appears therefore
much more natural to impose that {\em the classically induced} reality
conditions {\em on basic observables} should become adjointness-conditions
with respect
to the scalar product (roughly speaking one uses the observables as the
basic
coordinates of the theory without caring about the 'substructure' in terms
of the gauge-variant variables $A_a^i$ and $\tilde{E}^a_i$ i.e., if
$O=\bar{O}$ is a classical observable, then we want
to fix the scalar product by $O^\dagger=O$ whatever the expression of O in
terms
of $A_a^i$ and $\tilde{E}^a_i$ might be). This requirement is reasonable
due to the following argument : \\
The scalar product should integrate only over those coordinates on
which physical states depend. But then the observables are anyway the only
operators for which the scalar product is defined because the observables
are
precisely those operators, which leave ${\cal H}_{phys}$ invariant.
So there is then in general no necessity and no possibility any more to
implement
the conditions (4.3) and (4.4).\\
This then has the consequence that one has to get rid of
the 'gauge group volume' of the naively defined scalar product, just as in
the usual Yang-Mills field theories. Since the gauge 'group' of general
relativity is no Lie group, the Faddeev-Popov (FP) procedure will have to be
substituted by its extension, the Batalin-Vilkovisky-Fradkin (BFV) method
(see ref. \cite{8}). \\
We will apply this idea with success  to the present model:

\subsection{Step 2 : Choice of a representation of ${\cal A}$ on a linear
space}

We choose the self-dual (Schrdinger-) representation :
\begin{eqnarray}
\hat{A}_I(x) \Psi=A_I(x)\Psi , \\
\hat{E}^I(x)\Psi=\frac{\delta}{\delta A_I(x)}\Psi \; ,
\end{eqnarray}
where the  up to now arbitrary holomorphic functionals
$\Psi=\Psi[A_I]$ of
the configuration variables $A_I$ form a linear space $ \cal H $
(in the language of geometric quantization
(see ref. \cite{11}) the self-dual representation corresponds to the
polarization
${\cal P}=\mbox{span}(\delta/\delta E^I)$). \\
The triad representation on the other hand is given by
\begin{eqnarray}
\hat{A}_I(x)\tilde{\Psi} & = &-\frac{\delta}{\delta E^I(x)}\tilde{\Psi} \\
\hat{E}^I(x)\tilde{\Psi} & = & E^I(x)\tilde{\Psi} \; ,
\end{eqnarray}
where  $\tilde{\Psi}=\tilde{\Psi}[E^I]$ is an up to now arbitrary holomorphic
functional of the momentum variables $E^I$.\\
In what follows, we will drop the hat again, which indicated an operator in
this subsection.

\subsection{Step 3 : Determination of the physical quantities}

\subsubsection{The physical subspace ${\cal H}_{phys}$}

Following  Dirac's prescription (\cite{12}) a state functional $\Psi$ belongs
to ${\cal H}_{phys}$ iff it is annihilated by the (smeared) operator-valued
distributions that correspond to the classical constraint functionals. Now
this is quite an ambiguous definition because an infinite number of
inequivalent
operators have the same classical limit, which is defined to be the
functional
that one obtains when turning back the operators into commuting functions
(this is the so-called operator-ordering problem). However, there exists a
selection principle :
\begin{definition}
i) An ordering of a set of first class constraint operators is called
consistent iff its algebra is (weakly) closed.
ii) An operator is called a quantum observable iff it (weakly) commutes with
all constraints.
\end{definition}
(Note that this definition does not imply that the classical counterpart of
an observable is real!)\\
If no consistent ordering exists, then the theory simply cannot be quantized
in the given representation. One can show that the set of quantum observables
is identical with the set of operators that leave ${\cal H}_{phys}$
invariant.\\
These and further results are proven, for example, in ref. \cite{12}.\\
The constraint operators consist of polynomials of basic operators which are
located at the same point x in $\Sigma^1$, so they are a priori ill-defined
and must first be regularized. However, regularizations via point-splitting
suggested for the self-dual representation of quantum gravity in the
literature
poses serious problems (see refs. \cite{13},\cite{14}). Its most naive
form (that is simply separating the points x occuring in the operator product
by an amount of $\epsilon$ which one sets equal to zero after all commutators
and actions on states are carried out) turns all operators in principle into
commuting objects, so that one is effectively dealing with a semi-classical
theory. This is one of the reasons, why the self-dual representation is
hard to make consistent (in contrast to the loop representation). \\
We will now show that \\
1.\ for our model consistent operator-orderings exist that {\em need no
regularization},\\
2.\ the choice of ordering has physical significance.\\
In order to do that, we begin with a {\em classical} analysis of the
structure of the
constraint functionals. This provides us with a reformulation of the
constraints which are at most linear in the momenta, a property which
allows the quantum constraints of  our model to be solved: \\ \\ {\bf
Classical analysis of
 the constraints} \newline
There is a qualitative difference between the cases $E \neq 0$ and $E =0$
(see eqs. (2.20), (2.21)) the first one is associated with non-degenerate
the second one with degenerate\cite{15} solutions. \\ {\bf Sector I: $E
\neq 0$,
 nondegenerate
metrics } \newline   We first dispose of the
trivial case $B^1=0$: \\ As $E\neq 0$ it follows from
the vector and scalar constraints (2.8) and (2.9) that $B^2 = B^3 =0$ if
$E^1 \neq 0$. This implies $\gamma:= A_1 + \alpha' =
(A_2 B^2 + A_3 B^3)/A =0$.
It is not difficult to show that the canonical coordinates $ \alpha$ (see eq.
(2.20)) $\gamma$ and $B^1$ have the canonical momenta $ {\cal G}, E^1 $ and
$ \sqrt{E/2} \cos (\alpha-\beta)$ respectively. \\ Now we have 3 (gauge)
conditions
$ (\gamma =0, B^1= 0, {\cal G}=0) $ for the above 6 canonical momenta. This
implies that the physical phase  space consists of just {\em one} point, as
can be seen as follows: \\ If $q(x), p(x)$ form a conjugate pair
 and $\lambda(x)$ is the Lagrangean multiplier
for the gauge
condition $p(x) =0$    then
it follows from $ \{q(x), \int_{\Sigma} dy \lambda(y)p(y)\} = \lambda(x) $
that
$p(x)$ generates just {\em one} gauge orbit with respect to $q$, a
representative
of which is $q=0$. As $p=0$ we see the validity of the assertion. \\ The
argument
can be generalized  immediately to our case by choosing an appropriate
polarization
of the phase space in which $ \gamma, B^1 $ and $ {\cal G} $ are the
canonical momenta. \\
The operator constraint method on the other hand
shows  that ${\cal H}_{phys}$ is the linear span of the single functional
$\delta[A-2]\delta[\gamma]$ where $\delta[\gamma]:=\prod_{x\in\Sigma}
\delta(\gamma(x))$
and is therefore isomorphic to C (complex numbers). The quantum theory is
thus also trivial: observables are only the constant functionals, the
natural scalar product is given by $<\Psi_1|\Psi_2>=\bar{\Psi}_1\Psi_2$. \\
If $B^1 =0$ and $ E^1 =0$ we no longer can conclude that $B^2=B^3=0$.
However, we have the same situation as above, the only difference being the
interchange of the roles of $\gamma$ and $E^1$. Note that the metric (2.21)
becomes highly singular if $E^1=0$! \\
We do not consider this trivial sectors any further.\\
We now turn to the main case ($E\neq, B^1 \neq 0 $) which
we call "sector I" and which corresponds to metrics that are degenerate at
most
at isolated points. \\
Taking suitable linear combinations of the constraint functions yields
\begin{eqnarray}
E^2 H-E^3 E^1 H_x=E(B^2 E^1+\frac{1}{2}B^1 E^2), \\
E^3 H+E^2 E^1 H_x=E(B^3 E^1+\frac{1}{2}B^1 E^3) \; .
\end{eqnarray}
So, for sector I the constraint functions $H$ and $H_x$ are classically
equivalent to the constraint
functions
\begin{eqnarray}
\phi_2:=E^2+\frac{2B^2}{B^1} E^1 \; , \\
\phi_3:=E^3+\frac{2B^3}{B^1} E^1 \; ,
\end{eqnarray}
because conversely
\begin{eqnarray}
\frac{B^1}{2}(E^2\phi_2+E^3\phi_3)=H, \\
-B^3\phi_2+B^2\phi_3=H_x \; .
\end{eqnarray}
It is obvious  that the assumption $B^1\not=0$ is needed for the
eqs. (4.11)-(4.14) to make sense.\\
Finally we conclude from the Gauss constraint that
\begin{eqnarray}
(B^1)^2{\cal G} & = & (B^1)^2(E^1)'+(B^1)^2 A_2(\phi_2-2B^2 E^1)-B^1 A_3
(\phi_3-2B^3 E^1)\nonumber \\
& = & ((B^1)^2(E^1)'+((B^1)^2)'E^1)+(B^1)^2(A_2\phi_2- A_3\phi_3)\nonumber\\
& = & ((B^1)^2 E^1)'+(B^1)^2(A_2\phi_2- A_3\phi_3) \; .
\end{eqnarray}
The relations (4.13), (4.14) and (4.16) imply that the constraint functions
$H,H_x$ and $\cal G$ are then equivalent
to the set $\phi_I,I=1,2,3$, where
\begin{equation} \phi_1=B^1(E^1)'+2(B^1)'E^1 \; . \end{equation}
On the constraint surface the eqs. $\phi_I=0$ can be integrated to give
\begin{eqnarray}
E^1 & = & \frac{c}{(B^1)^2},     \\
E^2 & = & -\frac{2c}{(B^1)^3}B^2,   \\
E^3 & = & -\frac{2c}{(B^1)^3}B^3  \; ,
\end{eqnarray}
where c is up to now an arbitrary complex function of time. \\
Note :\\
The above equations (4.17)-(4.19) can also be obtained by the methods of
Capovilla et al.\cite{16}. They are equivalent to the following (inverse)
'CDJ-matrix':
\begin{equation} (\Psi_{ij})^{-1}=\frac{c}{(B^1)^3}\mbox{diag}(1,-2,-2)\;.
\end{equation}
Compared to the CDJ-approach we are able to solve the Gauss
constraint of our model, too, because it is rather simple.
\\ {\bf Sector II: $E=0$, degenerate metrics} \newline
For sector II on the other hand it follows from
the reality conditions (4.3) and (4.4) and Gauss' law (2.7) that we have
on the constraint surface :
$(E^1)'=E^2=E^3=0\Rightarrow E^1=c$,
c some real (due to the reality conditions) function of t, while the
$A_I$ remain undetermined. This corresponds to a globally degenerate metric
(see formula (1.11)). Hence the constraint functions $H,H_x$ and $\cal G$ are
here equivalent to the set
\begin{equation} \chi_1=(E^1)',\chi_2=E^2,\chi_3=E^3. \end{equation}
This is a remarkable result : The constraint surface splits into precisely
2 disjoint nontrivial pieces  (which is  associated with the the fact that
2 Riemannian manifolds, one of which  possesses a globally degenerate metric
while the other one is degenerate at most at isolated points, cannot be
diffeomorphic
if   the vector constraint functional {\em does} generate diffeomorphisms
 because
one has to observe that the associated test functions (see eqs.\ (3.9),
(3.10))
cannot be chosen arbitrarily but in particular have to be smooth).\\
Furthermore, the constraints that restrict the phase space to these sectors
can be chosen {\em at most linear in the momenta} $E^I$. On each sector
the original set of
constraint functions $H,H_x,\cal G$, henceforth called 'the canonical set',
is {\em equivalent} to the new set $\phi_I (\chi_I)$, henceforth called
the BRST-set
(this name will be justified in section (4.4)), where
we {\em define 2 sets of constraint functionals to be equivalent iff they
restrict the phase space to the same constraint surface}.
\\  \\ {\bf Quantum analysis of the constraints}\\ Let us
turn now to the quantum theory and compute the physical subspace with the
help of operator constraints.
The basic idea is to use the
constraint functions $\phi_I$  and $\chi_I$ instead of the original
ones. This is justified by showing that there is a regular
(operator-valued) transformation
between the 2 sets on the operator level. We refer to the appendix
for the proof.\\For the moment, consider the set of equations
(4.17)-(4.19)). Translating
these into quantum theory, one obtains simple functional differential
equations of the type \begin{equation} \frac{\delta F}{\delta
A_I(x)}=f^I(A_J)(x) \; ,
\end{equation}which means heuristically that the (I,x)-th partial
derivative of the functional F is prescribed. The existence of a
solution (a 'potential' for
the 'vector field' $(f^I(A_J)(x))_{I=1..3,x\in \Sigma}$) of this problem
is guaranteed at least locally, iff the integrability conditions are
satisfied. But
for this model one can apply the following theorem:
\begin{theorem}Provided that one is given a field theory such
that \\1) the constraint algebra is of 1st
class, \\2) the number of configuration space variables coincides
with the number of constraints (3 per spatial point x in this model), then
the integrability
conditions are automatically satisfied. \end{theorem}Proof :\\Let $E^I(x)
:=f^I[A_J,x)$ (the notation means that $f^I$ may be a functional of $A_J$
and a
function of x) an arbitrary solution of the classical constraints
$\phi_K=\phi_K[A_J,E^I]=0$, where one needs only to know that
their Poisson-algebra weakly
closes. Let $\eta:=M^I\circ\phi_I,\zeta:=N^I\circ\phi_I\;
(M^I,N^I$test functions of, say, compact support). From $\eta[A_K,E^L=f^L]
\equiv 0$ we
have\begin{eqnarray}0 & = & \frac{\delta}{\delta A_I(x)}\eta[A_K,E^L=f^L]
 \nonumber \\  & = & (\frac{\delta}{\delta A_I(x)}\eta[A_K,E^L])_{|E^P=f^P}
+\int dy
(\frac{\delta}{\delta E^J(y)}\eta[A_K,E^L])_{E^P=f^P}\frac{\delta f^J(y)}{
\delta A_I(x)}.\end{eqnarray}Therefore\begin{eqnarray} & &
(\frac{\delta\eta}{\delta A_I(x)})_{|E^P=f^P}  =  -(\frac{\delta\eta}{
\delta E^J})_{E^P=f^P}\circ\frac{\delta f^J}{\delta A_I(x)}  \\
\mbox{and}~~~~~~~ &   &
(\frac{\delta\eta}{\delta A_I})_{|E^P=f^P}\circ(\frac{\delta\zeta}{
\delta E^I})_{|E^P=f^P}  \nonumber \\           & = & -\int_\Sigma
dx\int_\Sigma dy
(\frac{\delta\zeta}{\delta E^I(x)})_{|E^P=f^P} (\frac{\delta\eta}{
\delta E^J(y)})_{E^P=f^P}\frac{\delta f^J(y)}{\delta A_I(x)} \; .
\nonumber\end{eqnarray}Exchanging $\eta$ and $\zeta$ as well as the
summation-indices, $(x,I)\leftrightarrow (y,J)$, and finally using the
definition of the
Poisson-bracket yields\begin{equation}(\{\eta,\zeta\})_{E^P=f^P}=-i
\int_\Sigma dx\int_\Sigma dy (\frac{\delta\eta}{\delta E^I(x)})_{
|E^P=f^P}
(\frac{\delta\zeta}{\delta E^J(y)})_{E^P=f^P} [\frac{\delta f^J(y)}{
\delta A_I(x)}-\frac{\delta f^I(x)}{\delta A_J(y)}] \; .\end{equation}
Now the 1st
hypothesis that the constraint algebra closes weakly comes in,
i.e. there exist structure functions U, such that :
$\{\eta,\zeta\}=U[M^I,N^J]_{IJ}\;^K\circ\phi_K$, where the
following notation has been used :\begin{eqnarray} & &
\{M\ast\phi_I,N\ast\phi_J\}:=U[M,N]_{IJ}\;^K\circ\phi_K \nonumber \\ & &
:=\int_{\Sigma}dx\int_{\Sigma}dy\int_{\Sigma}dz U(x,y;z)_{IJ}\;^K M(x) N(y)
\phi_K(z)
\; .\end{eqnarray}Hence \begin{equation} \{\eta,\zeta\}_{E^I=f^I}=0 \; .
\end{equation}Finally, exploiting the 2nd hypothesis, one realizes that
the
functions\begin{equation} \tilde{M}^I:=(\frac{\delta\eta}{\delta E^I(x)})_{
|E^J=f^J},\tilde{N}^I:=(\frac{\delta\zeta}{\delta E^I(x)})_{|E^J=f^J} \; ,
\end{equation}due to the arbitrariness of the test functions $M^I,N^I$,
 can be varied independently of the given field distribution of the $A_I$,
as long as
they do not vanish identically (which will not be the case, for example,
if the constraints are at most linear and homogenous in momenta). Thus
they can be
interpreted as new test functions and the validity of the integrability
conditions\begin{equation}\int dx\int dy \tilde{M}^I(x)\tilde{N}^J(y)[
\frac{\delta
f^J(y)}{\delta A_I(x)}-\frac{\delta f^I(x)}{\delta A_J(y)}]=0\; \forall\;
\tilde{M}^I,\tilde{N}^J,A_K\end{equation}is shown. $\Box$\\ \\The
'potential' can now
be computed by generalizing the potential-formula for an integrable
vector field on finite dimensional manifolds to infinite dimensional
ones (written down for
a field theory based on the configuration variable $\varphi$):
\begin{equation}F[\varphi]-F[\varphi_0]=\int dx (\varphi(x)-\varphi_0(x))
\int_0^1
dt\frac{\delta F}{\delta\varphi(x)}_{|\varphi_t} \; ,\end{equation}where
$\varphi_t:=\varphi_0+t(\varphi-\varphi_0)$ and $\varphi_0$ is some fixed
field
configuration. \\ \\In order to apply this theorem to our model one only
has to check that the new set of constraints $\phi_I$ again has a closed
Poisson
algebra. This is done in the appendix. \\In our model the
general solution of the quantum constraints  $\lambda^I\circ\phi_I\Psi=0,\;
\lambda^I$ suitable test
functions, can therefore be written down at once. For simplicity
we choose $\varphi_0=0$, which is allowed because the zero-field
distribution is an element of
the phase space; the  constant $\Psi[\varphi_0]$ may be neglected,
because the constant state functional always solves the constraints in
Ashtekar's formulation of general relativity. \\ \\{\bf  Sector I} :
\\Let $F:=\ln(\Psi_I)$, then\begin{eqnarray}F[A_I] & = & \int_{\Sigma}dx
A_I(x)\int_0^1
dt(\frac{\delta F}{\delta A_I(x)})_{|t A_I} \nonumber
\\ & = & c\int_{\Sigma}dx\int_0^1 dt[A^1 f(t A_I)+2 A^2(B^2\dot{f})_{
|t A_I}+2 A^3(B^3\dot{f})_{|t
A_I}](x) \nonumber \\ & = & c\int_{\Sigma}dx\int_0^1 dt[A^1 f(t^2 A)+t[2t A
\dot{f}(t^2 A)]A_1+[2t A\dot{f}(t^2 A)]\alpha'](x) \nonumber \\ & = &
c\int_{\Sigma}dx[(A^1+\alpha') f(t^2 A)]_0^1(x)   =   c\int_{\Sigma}dx(
\gamma f)(x)-\frac{c}{2}\int_{\Sigma}dx\alpha'(x)  \nonumber \\ & \iff &
\Psi_I[A_J]=\exp(c\int_{\Sigma}dx(\gamma f)(x)) \; ,\end{eqnarray}
where we used the abbreviations $\gamma =A_1+\alpha',f:=1/(B^1)^2$ and the
integral over the
total differential was neglected because it is a constant:
$\alpha\to\lim_{r\to\infty} \arctan(r^{1+\epsilon}\sqrt{2}/c_2)$.
\\We must show now that the
integrand of the functional F is not divergent at infinity :\\
We have $A_1+\alpha'=(A_2 B^2+A_3 B^3)/A$. The fall-off of the magnetic
fields is given by (see
section 3)\begin{eqnarray}B^1 & = & \frac{1}{2}((A_2)^2+(A_3)^2-2)\to\frac{
(a_3)^2}{r^2}+\frac{a_3\sqrt{2}}{r}+O(r^{-2-\epsilon})=O(r^{-1}),  \\B^2 &
= &
A_3'+A_1 A_2\to-\frac{a_3}{r^2}+O(r^{-3-\epsilon})=O(r^{-2}), \nonumber \\
B^3 & = & -A_2'+A_1
A_3\to\frac{c_2+c_1\sqrt{2}}{r^{2+\epsilon}}+O(r^{-3-\epsilon})=O(r^{-3-
\epsilon}) \; , \nonumber\end{eqnarray}because we imposed $\sqrt{2}c_1+c_2
=0$,
compare (3.12), whence\begin{equation}\frac{\gamma}{(B^1)^2}=O(r^{-1-
\epsilon}) \; .\end{equation}Finally one has to show that no surface term
survives when
acting with the (smeared) constraint-operators on F(note that the fall-off
of the associated test-functions corresponds to a {\em proper} gauge
transformation
because the Dirac condition only says that a physical state shall be
gauge invariant. It needs not be invariant under a symmetry transformation,
which
corresponds to non-vanishing boundary terms of the action functional).
This is equivalent to the proof that with our choice of function space
to which the
connections belong the functional F is indeed (functionally) differentiable.
 By definition, the functional derivative of a functional $f=f[\varphi]$
(if it
exists) is given by\begin{equation}\int_\Sigma dx T(x)\frac{\delta f}{
\delta\varphi(x)}:=\lim_{t\to 0}\frac{1}{t}(f[\varphi+t T]-f[\varphi]) \;
,\end{equation}where the test function $T=:\delta\varphi$, the variation
of $\varphi$, must fit into the function space of field distributions
$\varphi$.
Moreover, the so defined function $\delta f/\delta\varphi(x)$ must
belong to the allowed set functions of $\varphi$ (in our case, this set
consists of smooth
functions on the phase space).\\Since $(A_1)'$ does not appear in F we
need only inspect the surface terms that arise when varying with respect
to
$A_2,A_3$. We have\begin{eqnarray}& & \lim_{t\to 0}\frac{1}{t}(F[A_I+t
\delta A_2]-F[A_I])_{surface\;term}=-\int_{\partial\Sigma}\frac{A_3}{A
(B^1)^2}\delta
A_2,  \\& & \lim_{t\to 0}\frac{1}{t}(F[A_I+t\delta A_3]-F[A_I])_{surface\;
term}=+\int_{\partial\Sigma}\frac{A_2}{A (B^1)^2}\delta A_3 \; .
\nonumber\end{eqnarray}Inspecting formula (3.7) one can easily see that
the fall-off defined there suffices to make the surface terms vanish.\\ \\
{\bf Sector
II} :\\Let $G:=\ln(\Psi_{II})$\begin{eqnarray}G[A_I] & = & \int_{\Sigma}dx
A_I(x)\int_0^1 dt(\frac{\delta F}{\delta A_I(x)})_{|t A_I}        =
 c\int_{\Sigma}dx A_1(x)\int_0^1 dt        =  c\int_{\Sigma}dx A_1(x)
\nonumber \\    & \iff & \Psi_{II}=\exp(c\int_{\Sigma}dx A_1(x)) \; .
\end{eqnarray}The
functional G is well defined because, according to the choice of function
space (=phase space) to which the connections belong, the integrand falls
off at
infinity as $O(r^{-2-\epsilon})$.\\ \par Thus, we have found the general
solution of the quantum constraints : \\${\cal H}_{phys}$ consists of 2
sectors. On
sector I physical states are yet arbitrary functions of the functional F,
$\Psi[A_I]=\Psi_I(F)$, on sector II physical states are yet arbitrary
functions of
the functional G,$\Psi[A_I]=\Psi_{II}(G)$.

It is now of considerable interest to see whether the states just computed
are in fact also annihilated by the constraint operators of the canonical
set.
This is however obvious, since the constraint operators of the canonical set
can be ordered in such a way that they are just linear
combinations with operator-valued coefficients of the BRST-set (D is
{\em any}
function of the connections and its spatial derivatives) :
\begin{eqnarray}
{\cal G} & = & \frac{1}{B^1}\phi_1+A_2\phi_3-A_3\phi_2,  \\
H_x & = & B^2\phi_3-B^3\phi_2, \nonumber \\
H & = & \frac{1}{2 D}(E^2 B^1 D\phi_2+E^3 B^1 D\phi_3) \; , \nonumber
\end{eqnarray}
where the constraints of the BRST-set {\em appear always on the rhs}.
\\ An interesting choice is given by
\begin{equation}
D=(B^1 B)^{-1/2},B=(B^2)^2+(B^3)^2 \; ,
\end{equation}
because then the scalar constraint functional takes the form of a
d'Alembert-Beltrami functional differential operator :
\begin{equation} H=\frac{1}{2}\frac{1}{\sqrt{-\det(G)}}E^I G_{IJ}
\sqrt{-\det(G)} E^J \; ,\end{equation}
with respect to the 'magnetic metric'
\begin{equation} (G_{IJ})= \left ( \begin{array}{*{2}{c@{\;}}c} 0
& B^2 & B^3 \\
B^2 & B^1 & 0 \\ B^3 & 0 & B^1 \end{array} \right ) \; , \end{equation}
with determinant $-\det(G)=-\det((G^{IJ}))=(B^1 B)^{-1}$.\\
\\
We now show that the choice of ordering indeed has physical significance: \\
Namely, the physical states that correspond to sector II
of the constraint surface ($G=\int_\Sigma dx A_1$)
are in fact {\em not annihilated} by the scalar constraint operator ordered
as above for a general choice of D. For D=1, for example, we get :
\begin{equation}
\tiN\circ H\Psi_{II}(G)=\int_\Sigma dx\tiN(E^2 B^2+E^3 B^3)\dot{\Psi}_{II}
=\dot{\Psi}_{II}\delta(0)\int_\Sigma dx\tiN A_1=\infty\not=0 \; .
\end{equation}
Although there is at least one choice for D such that
the scalar constraint annihilates the states of both sectors, namely $D=1/B$
(which may be checked by calculations similar to those above), we were not
able to find {\em observables} with respect to a choice of D that are
well-defined on both sectors simultanously (a proof that $D=1/B$ fails to
fulfil this requirement will be given in the next subsection.
In case that such a choice of D would exist we would have a nice selection
principle for D at our disposal, but we suspect that no such D can be found
and conjecture that the following 'superselection rule' holds : \\
It is impossible to superpose states belonging to different sectors, because
they either fail to be both annihilated by the same ordering of the scalar
constraint or no observables common to both sectors do exist.\\
This corresponds to the classical situation that {\em the constraint surface
is unconnected}. Thus, the ordering has physical relevance and it is an
important structural element of quantum theory.
Furthermore, one can argue that at least a regularization of the scalar
constraint by the naive version of point-splitting mentioned in section
(4.3.1), would have given a wrong
result, because it would not differentiate between the 2 sectors.\\
It should be clear by now that the ordering has great influence on the
resulting quantum theory. In particular if the above mentioned function D
would exist the number
of degrees of freedom would double (see next subsection), because physical
states would depend on F and G, which were then both observables.\\ \\
At the end of this subsection we display, for completeness sake, an operator
ordering of
the canonical constraints that is appropriate for sector II :
\begin{equation}
H=\frac{1}{2}((2 B^2 E^1+B^1 E^2)E^2+(2 B^3 E^1+B^1 E^2)E^3) \; .
\end{equation}
The ordering of the kinematical constraints remains the same.

\subsubsection{Observables and reality conditions}

{}From now on we deal with sector I only.\\
As already stated at the beginning of this section, (quantum) observables are
exactly those operators that leave the physical subspace invariant. As the
latter
here consists of yet arbitrary, complex functions of F, an observable O is an
operator
which maps a function $\Psi=\Psi(F)$ of F into a new function $\Phi=\Phi(F)$.
We define the dense subspace $\cal P$ of ${\cal H}_{phys}$ with respect to,
say, the supremum-topology (or any finer one) with F restricted to a bounded
domain in C, to be the polynomials of F of finite degree.
Consider then the class ${\cal P}_n$ of monomials of a given degree n.
An observable maps these classes into each other or finite sums thereof. All
observables are
known if one knows any one observable that changes the degree n by +1,-1,0.
These are precisely the multiplication operator F, the derivation operator
$\partial/\partial F$ and the unit operator 1, which are the basic
observables.
Any other lies in the algebra generated by these 3 operators. They are
densely defined (on $\cal P$).\\
Let us represent these operators in terms of Ashtekar's variables. For
F this is already done :
\begin{equation} \hat{Q}:=-i\int_\Sigma dx(A_1+\alpha')(B^1)^{-2}=-iF \; ,
\end{equation}
where the motivation for the factor i will become clear in a moment.\\ For
the derivative operator a convenient choice is
\begin{equation} \hat{P}:=\int_\Sigma dx T(x)(B^1(x))^2 E^1(x) \; .
\end{equation}
Here T is any smooth test function normalized to 1,
\begin{equation} \int_\Sigma dx T(x)=1 \; ,\end{equation}
introduced in order that the basic operators $\hat{Q}$  and $\hat{P}$
become conjugate :
\begin{equation} [\hat{Q},\hat{P}]=i \; . \end{equation}
Then for any function $\Psi$ of $Q=-i F$ :
\begin{equation} \hat{Q}\Psi(Q)=Q\Psi(Q),\hat{P}\Psi(Q)=-i\frac{\partial}
{\partial Q}\Psi(Q) \; .\end{equation}
Although the theorem mentioned after definition (4.1) ensures that the
observables $\hat{Q}$ and $\hat{P}$ in fact
weakly commute with the constraint operators, an explicit proof will be
given in the appendix.\\
Now it is easy to show that these observables are in fact no observables on
both sectors for the choice $D=1/B$ mentioned in the last subsection,
because the product functional (taking $\Psi=\int_\Sigma dx A_1=:G$ as a
physical state)
\begin{equation} i \hat{Q} G=F G=\int_\Sigma dx(\frac{A_1+\alpha'}{(B^1)^2})
(x)\int_\Sigma dy A_1(y), \end{equation}
which should be a physical state again, is not annihilated by the scalar
constraint.
In fact, a straightforward but rather involved computation reveals that
\begin{equation}
\tiN\circ\hat{H} F G =-2\int_\Sigma dx\tiN\frac{B}{(B^1)^3}\not=0 \; .
\end{equation}
Since the 'multiplication-operator' $\hat{Q}$ is a basic ingredient of
the quantum theory
it must not be neglected and thus we are {\em forced} to select a sector, in
order that $\hat{Q}$ is well-defined. This is the proof promised in the last
subsection.\\
\\
Let us now determine the {\em classical} reality conditions of the
observables
on the constraint surface. Since observables are invariant along the gauge
orbit, these reality conditions are then also valid on the reduced phase
space.
An essential  theorem is the following (which  holds for full gravity, too) :
\begin{theorem}
If O is an observable, then Re(O) and Im(O) are also seperately observables.
\end{theorem}
Proof :\\
One has to show that $\{\bar{O},\Phi\}\approx 0$ ($\approx 0$ means =0 on
the constraint surface) for all (smeared) linear combinations of the
{\em canonical} constraint functions $\Phi$. We have (indices are suppressed)
$\bar{O}(A,E):=O(\bar{A},\bar{E})=O(\bar{A},E)$. Viewing E and the
anti-self-dual
connection $\bar{A}$ as independent variables (where $\bar{A}=-A+2\Gamma(E)
\Rightarrow \delta/\delta\bar{A}=-\delta/\delta A$),
it follows from the definition of the Poisson bracket that
\begin{eqnarray}
0 & \approx & \overline{\{O,\Phi\}}=\overline{-i(\frac{\delta O}{\delta A}
\circ\frac{\delta\Phi}{\delta E}-\frac{\delta O}{\delta E}\circ
\frac{\delta\Phi}{\delta A})}
=  i(\frac{\delta\bar{O}}{\delta\bar{A}}\circ\frac{\delta\bar{\Phi}}
{\delta E}-\frac{\delta\bar{O}}{\delta E}\circ\frac{\delta\bar{\Phi}}
{\delta\bar{A}}) \nonumber \\
 & = & -i(\frac{\delta\bar{O}}{\delta A}\circ\frac{\delta\bar{\Phi}}
{delta E}-\frac{\delta\bar{O}}{\delta E}\circ\frac{\delta\bar{\Phi}}
{\delta A})  =  \{\bar{O},\bar{\Phi}\} \; .
\end{eqnarray}
But $\bar{\Phi}\approx 0$, just as in full gravity, as can easily be
shown.\\
$\Box$\\
Hence, without loss of generality one can always choose a set of basic
{\em real} observables
(note that 2 observables are identified, if they agree on the constraint
surface).
The next theorem, unfortunately, does not carry over to full gravity.
\begin{theorem}
The classical magnetic fields are (weakly) real.
\end{theorem}
Proof :\\
Using eqn. (2.23) we obtain
\begin{eqnarray}
\bar{A} & = & (A_2-2\Gamma_2)^2+(A_3-2\Gamma_3)^2
=  A-4(A_2\Gamma_2+A_3\Gamma_3)+4((\Gamma_2)^2+(\Gamma_3)^2) \nonumber \\
        & = & A+\frac{4 (E^1)'}{E}(A_2 E^3-A_3E^2)+\frac{4 ((E^1)')^2}{E}
         =  A+\frac{4 (E^1)'}{E}{\cal G} \approx A \; ,
\end{eqnarray}
hence $B^1$ is already real. Next we solve the constraints for $B^2,B^3$ :
\begin{eqnarray}
B^2 & = & \frac{E^3}{E}H_x+\frac{E^2}{E^1 E}H-\frac{E^2}{2 E^1}B^1 \\
B^3 & = & -\frac{E^2}{E}H_x+\frac{E^3}{E^1 E}H-\frac{E^3}{2 E^1}B^1 \; ,
\end{eqnarray}
to conclude by means of the preceding theorem and the (weak) reality of
$B^1$, that $\bar{B^2}\approx B^2,\bar{B^3}\approx B^3$.\\
$\Box$\\
With these results it is an easy task to show that the observables Q and P
are indeed (weakly) real, thereby justifying the factor i contained
in Q . It is already obvious that P is real while for Q
\begin{eqnarray}
\gamma & := & A_1+\alpha'=\frac{1}{A}(A_2 B^2+A_3 B^3) \Rightarrow \nonumber
\\
\bar{\gamma} & \approx & \frac{1}{A}((-A_2+2\Gamma_2)B^2+(-A_3+2\Gamma_3)B^3)
\nonumber \\
 & = & -\gamma+\frac{2(E^1)'}{A E}(B^2(-E^3)+B^3 E^2) \nonumber \\
 & = & -\gamma-\frac{2(E^1)'}{A E}H_x \approx  -\gamma \; ,
\end{eqnarray}
so that finally
\begin{equation} \bar{Q}=-\overline{i\int_{\Sigma}dx\gamma(B^1)^{-2}}\approx
Q  \; . \end{equation}
\\
These are the reality conditions for the {\em classical} basic observables.
Since all gauge invariant quantities can be
constructed from them, there are only 2 basic variables left on the reduced
phase space
which thus turns out to be 2-dimensional.\\
As proposed in section (4.3.1), the scalar product should simultanously turn
the (real) observables into self-adjoint operators on the physical
subspace, which then acquires the stucture of a Hilbert space.\\
Physical states depend only on the real observable Q, so the scalar product
should only integrate over Q.
One can thus already intuitively guess that the desired Hilbert-space is
given by ${\cal H}_{phys}=L_2(R,dQ)$. In the next subsection
 we will sketch the main ideas to
justify this more systematically. The details are given elsewhere
 (see ref. \cite{17}).

\subsection{Step 4 : Construction of the scalar product}

The motivation for the following construction is the expression for the
pre-scalar product
in the framework of geometric quantization (see, e.g., ref. \cite{2})
for the self-dual representation :
\begin{equation} <\Psi|\Phi>=\int_P[i dA_a^i\wedge d\tilde{E}^a_i]\exp(k)
\bar{\Psi}[A]\Phi[A] \; , \end{equation}
where k is some strictly real functional of $\tilde{E}^a_i$ and $A_a^i$,
which guarantees
convergence of the functional integral over the whole phase space P.
Geometric
quantization, unfortunately, does not show how this inner product has to be
modified when the integration extends only over the Lagrangean subspace
(see reference \cite{11})
determined by the choice of polarization, in order that it becomes a
scalar product.
Furthermore, no notice is taken of possible divergencies that appear
typically
in gauge theories when integrating along the fibres.\\
The latter observation suggests to look at the partition function of gauge
theories. Here one cures the problem by the Faddeev-Popov (FP) procedure,
or more generally if the gauge group is not a proper Lie group, by the
Batalin-Fradkin-Vilkovisky (BFV) method (\cite{8}) :
\begin{equation} Z[j]=\int_P d\mu_L \exp(-I[j]+\int_{t_1}^{t_2}dt
\{G,\Omega\}) \; , \end{equation}
where $I[j]$ is the Euclidean action, j the external current to generate the
n-point-Schwinger-functions, $\mu_L$ is the Liouville-measure on the
ghost-extended
phase space, $\Omega$ is the BRST-charge and G some gauge-fixing functional.
The important point is that one encounters precisely the same situation in
the case of the pre-scalar product : $I[0]$ and $\bar{\Psi}\Phi$ are both
gauge invariant, because physical states depend only on observables,
$\{G,\Omega\}$ and k are both gauge fixing exponents. But while the
BFV-theorem
guarantees that the partition function does not depend on G, this is not
obvious for the pre-scalar product.\\
The basic idea is now simply to replace the above exponent $k$  by an
analogous expression,
thereby introducing ghosts and to extend the BFV theorem to our case. If
it is
possible to integrate out the ghosts and the momenta then this
will provide us with an elegant method to obtain a gauge invariant inner
product
that integrates over the configuration coordinates of the reduced phase
space only.
As proved in the last subsection these can always chosen to be real because
they are observables and therefore it should
be possible to turn their quantum version into self-adjoint operators on
${\cal H}_{phys}$
simply by integrating over the real line. Thus, it will be not necessary to
introduce a complicated measure on the full phase space in order to raise the
reality conditions on $A_a^i$ and $\tilde{E}^a_i$ to the operator level
which, as already
argued in section (4.1.1), will in general not even guarantee the observables
to become self-adjoint operators.\\
We will apply the above idea in the following, which is, of course, only
practicable
because we know all the observables.\\
\\
The proposal for the scalar product in full pure quantum gravity is then
\begin{equation} <\Psi|\Phi>=\frac{1}{N}\int_P[i dA_a^i\wedge d\tilde{E}^a_i
\wedge dc^a_i\wedge d\rho_a^i]\exp(\{G,\Omega\})\bar{\Psi}[\check{A}]\Phi
[\check{A}] \; , \end{equation}
where N is some yet unknown normalization constant, $c^a_i$ and $\rho_a^i$
are the ghosts
and their conjugate momenta respectively and the accent circumflex appearing
in the argument
of the wave functional is to indicate that the observables on which the
latter depend have to be BRST-extended in general, in order that they
strongly commute
with the BRST charge $\Omega$ (which, accordintg to \cite{8}, is always
possible for any choice of a  set of
constraints). The proof that this scalar product is
in fact independent of the gauge fixing functional G follows along the chain
of arguments of the BFV-theorem (compare chapter 9 of reference \cite{8})
and will be omitted here.\\
Let us now apply this construction to our model :\\
The method is, of course, only of practical interest if it is possible to
carry out all ghost and momentum integrations. There are 2 means to achieve
this : \\
1) the choice of the set of constraints, \\
2) the choice of the gauge fixing functional.\\
One can show (compare \cite{8}) that the rank of the algebra as well as
the 'length' of the BRST-extension of observables depends on the choice of
the constraints. One should try to get these as small as possible.\\
As it is easy to integrate Gaussian bosonic integrals, one should
keep the momenta outside the structure functions of the constraint algebra,
since otherwise they would appear inside the ghost determinant after
having carried out the ghost integrations.
This latter condition turns out to be in accordance with the requirement to
have a short BRST-extension of the observables because then the extension
(Poisson-) commutes
with that part of $\Omega$ which is of higher order in the ghosts and
anti-ghosts.
Finally, the gauge fixing functional should also contain no momenta if
possible, for otherwise a quartic ghost term would in general survive
in $\{G,\Omega\}$ because
we argued that the structure functions should depend on the connections
only.\\
{\em Fortunately, the BRST-set of constraints in our model obeys all
these requirements},
thereby giving reason for its name. Namely, according to the appendix \\
1) the rank of the BRST-algebra is 1, \\
2) the structure functions do not involve momenta, \\
3) as the BRST-set is at most linear in the momenta the annihilation of
$Q$ by the (smeared)
constraints $\phi_I$ is equivalent to the (strong) commutativity of $Q$ with
$\phi_I$. As the structure functions do not depend on the momenta the
observable $Q$ is already BRST-closed, thus no BRST-extension occurs.\\
It then turns out that one can reduce the scalar product
\begin{equation} <\Psi|\Phi>=\frac{1}{N}\int_P[i dA_I\wedge dE^I\wedge dc^I
\wedge d\rho_I]\exp(\{G,\Omega\})\bar{\Psi}(Q)\Phi(Q) \end{equation}
to the standard $L_2(R,dQ)$-scalar-product
\begin{equation}
<\Psi|\Phi>:=\int_R dQ\bar{\Psi}(Q)\Phi(Q)
\end{equation}
simply by integrating out the undesired coordinates $c^I,\rho_I,E^I$ and the
continuous degrees of freedom contained in $A_I$, after imposing a suitably
chosen gauge-fixing functional G compatible with the definition of phase
space,
which is independent of the $E^I$ (a possible choice for G can be obtained by
requiring
 that the connection coefficients $A_I$  fall off
with a specific power at infinity, which is compatible with the requirements
 of section 3, see ref.\ \cite{17}).\\
For the sake of completeness  consider the complications we would have
encountered had we
insisted on implementing the adjointness-conditions (4.3) and (4.4). The
reader may
convince himself that one could not give any sense to $Q^\dagger$ in that
case because
the integrand of $Q$  depends non-analytically on the connections, not to
speak of
operator-ordering difficulties arising in the definition of an infinite
series of functional derivatives.

\section{Why two degrees of freedom ?}
The classical Birkhoff theorem states that any spherically symmetric
solution of the Einstein equations in vacuum is {\em gauge
equivalent to a Schwarzschild solution} expressed in its standard static
foliation (slicing of space and time),
 where 2 solutions $g,\bar{g}$ are
defined to be gauge-related if there exists a diffeomorphism $(M,g)\to
(M,\bar{g})$.
The Schwarzschild solutions are parametrized by {\em one} real parameter,
the
Schwarzschild mass m. Moreover, there exists a 1-1-correspondence between
the numbers
of gauge-inequivalent classical solutions and gauge-inequivalent Cauchy data
for (gauge) field theories such as general relativity that admit a
Hamiltonian formulation (see ref. \cite{18}). The latter, in turn, can be
identified with the number of degrees of freedom, i.e. the dimension of
the reduced phase space. Thus we seem to have a serious problem : How can
our analysis be reconciled with Birkhoff's theorem?\\
The answer is similar to that in ref. \cite{6} : the Lagrangean formulation
of a field theory on which the proof of  Birkhoff's theorem is based and
its corresponding
Hamiltonian formulation use different notions of gauge :\\
Roughly speaking, from the spacetime (Lagrangean) point of view one uses any
4-diffeomor\-phism
to relate metrics, where, for instance, it is not important whether the
diffeomorphism
diverges at infinity or not (see \cite{6}). On the other hand, from the
Hamiltonian point of
view only those 'diffeomorphisms' are allowed that 1) fit into the
definition
of phase space and 2) are generated by the constraints.\\
Now in our model, the observables P and Q (Poisson-) commute weakly
with all the
constraints which in particular means that they are gauge independent,
and thus {\em cannot} be gauged away. \\  We will now show that for any
solution of the classical equations of motion P is essentially just the
Schwarzschild mass squared,
while the observable Q can be gauged to zero from the spacetime point of
view .
The latter can be expected from the following expression for Q (obtained
by using the formulas (2.23),(2.24) and the relation $A_I=\Gamma_I+iK_I$):
\begin{equation}
Q=\frac{1}{4c}\int_\Sigma dx\frac{N^x}{\tiN}\frac{(E^1)'}{E^1(1+
\sqrt{\frac{E^1}{c}})}\; ,
\end{equation}
valid for those special field configurations for which the functions
$E^1,E,N^x$ and $\tiN$ are {\em stationary} . According to Birkhoff's
theorem
the shift-vector $N^x$ can be made to zero by a diffeomorphism. One
immediately
sees that one is dealing with non-static foliations ($N^x\not =0$) if
$Q\not =0$ in this special case.\\
The situation is thus as follows : We know that Q is gauge invariant from
the
Hamiltonian point of view but can be gauged to zero from the Lagrangean
point
of view. There are 2 possible resolutions of this apparent contradiction :\\
Either we have  $Q=0$ identically for any physical field configuration or
 the Hamiltonian and Lagrangean notions of gauge are here  different, too.\\
The subsequent discussion is intended to rule out the first option by
explicit calculations. It shows in particular
that the classical physical spectrum of Q is continuous (and not discrete as
one might expect) and that there is therefore a well-defined 2-dimensional
reduced phase space. We do this - for the sake of
simplicity - by considering {\em stationary}  solutions of the
equations of motion for sector I. The calculations reveal that Q=0 for all
 Schwarzschild
solutions in the standard static slicing,
 but that in general $Q \neq 0$, e.g. for  stationary slicings. \\
 \\
Inserting the solution (formulas (4.17)-(4.19)) of the constraint equations
into the
evolution equations (2.17) one can check that those for $E^I$ are identically
satisfied as they have to be, so that one ends up with
\begin{eqnarray}
\dot{A}_1 & = & \Lambda'-2i c\tiN\frac{B}{(B^1)^3},  \\
\dot{A}_2 & = & \Lambda A_3-N^x B^3-i c\tiN\frac{B^2}{(B^1)^2}, \nonumber \\
\dot{A}_3 & = & -\Lambda A_2+N^x B^2-i c\tiN\frac{B^3}{(B^1)^2} \; .
\nonumber
\end{eqnarray}
On the constraint surface the independent metric components read
\begin{equation} q_{\theta\theta}=\frac{c}{(B^1)^2},q_{xx}=\frac{2c B}
{(B^1)^4} \;.  \end{equation}
We will impose coordinate (gauge) conditions on $B^1,B$ later.\\
 Assuming
the system to be stationary ($\dot{A}_I=0$) but not necessarily static the
evolution equations can be uniquely solved for the multipliers :
\begin{eqnarray}
\Lambda & = & i k (B^1)^2\;,\; k=\mbox{const.}\; , \\
N^x & = & \frac{\Lambda}{B}\sqrt{2 B(1+B^1)-((B^1)')^2}\; , \nonumber \\
\tiN & = & \frac{(B^1)^3}{2ic B}\Lambda'=2 k\frac{(B^1)'}{q_{xx}} \; ,
\nonumber
\end{eqnarray}
while the reality conditions $Re(A_I)=\Gamma_I$ turn out to be equivalent
to imposing the reality of the magnetic fields because, remarkably enough,
the 'electric' spin connections on the constraint surface become
{\em precisely}
the 'magnetic' ones,
\begin{equation} (\Gamma_1,\Gamma_2,\Gamma_3)=(-\beta',-(B^1)'\frac{B^3}{B},
(B^1)'\frac{B^2}{B}) \end{equation}
(compare formula (2.23)). They have the general solution (the equations (5.2)
are not needed to derive the following relations) :
\begin{eqnarray}
(Im(A_2),Im(A_3)) & = & \rho\;(Re(A_3),-Re(A_2)),  \\
(Re(A_1),Im(A_1)) & = & ((\arctan(\frac{Re(A_3)}{Re(A_2)}))'\; ,\rho'+
\frac{\rho\;(A^R)'}{2 A^R}), \nonumber \\
\rho & := & \frac{1}{(B^1)'}\sqrt{2 B(1+B^1)-((B^1)')^2}, \nonumber \\
A^R & := & (Re(A_2))^2+(Re(A_3))^2=2\frac{1+B^1}{1-\rho^2}, \nonumber
\end{eqnarray}
which implies that all quantities may be expressed in terms of $B^1,B$
and $Re(A_1)$.
\\ On the constraint surface the observables are classically given by
\begin{equation} P=c,Q=i\int_\Sigma dx\frac{1}{2(1+B^1)(B^1)^2}
\sqrt{2B(B^1+1)-((B^1)')^2}:=\int_\Sigma dx h \; ,\end{equation}
and here one even does not have to know $Re(A_1)$. \\
We are now able to impose coordinate conditions compatible
with the definition of phase space. In view of formula (5.3) and
according to section 3 we make the following ansatz
\begin{equation}
q_{\theta\theta}=x^2\Rightarrow B^1=\pm \frac{\sqrt{c}}{x},q_{xx}=
(1+\sigma(x))^{-1} \; ,
\end{equation}
where $\sigma$ is assumed to possess a Laurent-expansion in x. It is,
however, not
arbitrary because the rest of the conditions derived in section 3 are not
satisfied yet. They ensured in particular that the integral over h in
equation (5.7) converges. Computing h for our ansatz yields
\begin{eqnarray}
h & = & i\frac{x^2}{2c}(1\pm\frac{\sqrt{c}}{x})^{-1}\sqrt{\frac{c}{x^4}
\frac{1\pm\frac{\sqrt{c}}{x}}{1+\sigma(x)}-\frac{c}{x^4}}
   =\pm i\frac{1}{2\sqrt{c}}(1\pm\frac{\sqrt{c}}{x})^{-1}\sqrt{\frac{\pm
   \frac{\sqrt{c}}{x}-\sigma(x)}{1+\sigma(x)}}  \\
  & \stackrel{!}{=} & O(r^{-1-\epsilon})\; (x\to r\to\infty)  \; . \nonumber
\end{eqnarray}
Thus, in order to ensure convergence of the integral over h we must require
$\sigma(x)=\pm\frac{\sqrt{c}}{x}+z(x),z(r)=O(r^{-3})$.
This shows that $\pm\sqrt{c}=-2 m$, where m is the Schwarzschild mass.
Since h is real we have to require that z is a strictly positive
function. Furthermore, recalling the fall-off of $B^1$, formula (4.32) and
comparing with formula (5.8) we deduce that $\sqrt{2}a_3=-2m$.\\
So far we have imposed coordinate conditions in the asymptotic region only
and are still free to join smoothly other conditions in other charts. But, in
order to
discuss a concrete example, let us assume that $\Sigma$ is topologically
$R^1$,
i.e. has 2 ends, and that it is possible to choose a global chart. Then a
choice that
makes h integrable in this chart and everywhere well-defined is, for
instance,
\begin{eqnarray}
z(x) & := & \frac{a}{x^4}(1\pm\frac{\sqrt{c}}{x})^3\Rightarrow \\
h(x) & = & \pm\frac{1}{2\sqrt{c}\sqrt{\frac{x^4}{a}+(1\pm
\frac{\sqrt{c}}{x})^2}} \; ,
\end{eqnarray}
$a\ge 0$ being a constant and for $a>0$ the observable $Q$  does not vanish,
while
for $a=0$ we arrive at the usual Schwarzschild configuration. One may easily
check that the metric  encounters the usual Schwarzschild-type coordinate
singularities and
that there is a singularity at x=0 of a scalar curvature-polynomial. We do
not
care about that here because the integral over the densitized curvature
polynomial
is not an observable of our theory.\\
$Q$ also cannot be gauged away by use of the
constraint-generators due to its very definition as an observable. Indeed,
the metric
that corresponds to this solution of the equations of motion is {\em not}
static
but only stationary, which is indicated by the non-vanishing of the
shift-vector $N^x$ :
\begin{equation} N^x=\pm 2k\sqrt{c}(1+\sigma(x))\frac{1\pm
\frac{\sqrt{c}}{x}}{\sqrt{\frac{x^4}{a}+(1\pm\frac{\sqrt{c}}{x})^2}} .
\end{equation}
Note that the fall-off behaviour of $N^x$ fits into the definition of phase
space of section 3 and that k has to be real. Obviously $N^x$ vanishes iff
Q vanishes. How does this come about ?\\
Usually, when deriving the
Schwarzschild metric and  Birkhoff's theorem, (see \cite{18})
one argues that $N^x$ can be made to vanish because 'one may choose the
coordinates
t,x arbitrarily in the 2-surface $\theta,\phi=$const.'. More precisely
this means
the following :\\
In the spacetime picture, where $g_{tt}=N^2-(N^x)^2q_{xx},g_{xt}=q_{xx}N^x,
g_{xx}=q_{xx}$
(recall formula (1.12)),
we can choose a new time function $t=t(x,\tau)$ such that for the metric
$\hat{g}$ diffeomorphic to $g$ the relation $\hat{g}_{x\tau}=0$ holds. This
implies
\begin{equation} \delta g_{xx}:=\hat{g}_{xx}-g_{xx}=g_{tx}^2/g_{tt},
\end{equation}
where we use the symbol $\delta$ to denote the difference between 2 field
configurations which are equivalent by an infinitesimal gauge transformation,
whereas we denote the (infinitesimal) difference between 2 solutions
$\tilde{g}$
and $g$ of the equations of motion by $\Delta g=\tilde{g}-g$.\\
Since (up to 4-diffeomorphisms) solutions of the Hamiltonian equations of
motion
also solve the Euler-Lagrange equations we may use the above solutions
$\tilde{g}_{xx}:=g_{xx}(a\not =0)$ and $g_{xx}:=g_{xx}(a=0)$ in the
Lagrangean
framework and find $\Delta g_{xx}=\delta g_{xx}$. This is of course expected,
since Einstein's equations are 4-diffeomorphism covariant.\\
We will now show that $\tilde{g}$ and $g$, however, are not gauge-equivalent
from the Hamiltonian
point of view:\\
Consider first any constrained field theory with canonical coordinates
$\varphi$,
Lagrange multipliers $\lambda$ and constraints $C=0$. In order to arrive at a
representative of an initial data set  a specific gauge one
has to impose a coordinate condition, $\chi=0$, which we also did in (5.8).
The consistency conditions that this gauge is preserved under time evolution
($\dot{\chi}=0$) lead to a unique initial data set  $(\varphi;\lambda)$.\\
In the Hamiltonian picture one regards the configurations $(\hat{\varphi};
\hat{\lambda})$
and $(\varphi;\lambda)$ as infinitesimally gauge-related iff the
infinitesimal
gauge transformation effected by $\delta\lambda=\hat{\lambda}-\lambda$ solves
$\delta\varphi:=\{\varphi,\delta\lambda\circ C\}$, i.e. {\em iff the
transformation
is generated by the constraints}. Hence the necessary and sufficient
condition
for 2 initial data set  $(\tilde{\varphi};\tilde{\lambda})$ and $(\varphi;
\lambda)$
to be gauge equivalent, $\Delta\varphi=\delta\varphi$, in the Hamiltonian
picture
takes the form $\Delta\varphi\stackrel{!}{=}\delta\varphi:=\{\varphi,
\Delta\lambda\circ C\}$.\\
In our case, we have  $(\varphi;\lambda)\rightarrow(g;N^x,\tiN)$
and are dealing with the above solutions of the equations of motion that
correspond to vanishing or non-vanishing parameter $a$ respectively.
According to formula (5.4) we have
\begin{eqnarray}
 & & \Delta N^x=N^x(a)-N^x(a=0)=N^x(a)=O(r^{-2}),(\delta N^x)'=O(r^{-3}),
 \nonumber \\
 & & \Delta\tiN=\tiN(a)-\tiN(a=0)=O(r^{-6}) \; ,
\end{eqnarray}
so that the induced change of the metric components is given by
\begin{equation} \delta q_{xx}=\frac{1}{E^1}([\frac{1}{2}(\Delta N^x)E'+
(\Delta N^x)'E-(\Delta N^x)q_{xx}(E^1)']-i[(\Delta\tiN E E^1(A_1
-\Gamma_1)]) \; ,
\end{equation}
where the 1st bracket is $O(r^{-3})$, the 2nd $O(r^{-6})$, whereas
$\Delta q_{xx}=O(r^{-4})$ (see eqn. (5.10)).\\
This is the resolution of the apparent contradiction : the Hamiltonian
and the Lagrangean picture simply have different understandings of 'gauge', a
conclusion first drawn in ref. \cite{6}.\\
This is not unexpected : from the above
analysis it is obvious that in the Lagrangean and Hamiltonian picture
respectively, the change of $q_{xx}$ involves {\em different}
powers of $\Delta N^x$, namely  $\delta q_{xx}\propto(\Delta N^x)^2$ and
$\delta q_{xx}\propto\Delta N^x$ respectively (see eqs. (5.13) and  (5.15)).
\\
\\
It is worthwhile to compare the functional Q with the Chern-Simons
functional $C_2$
for the spherically symmetric case : Up to a boundary term which vanishes
with
our choice of phase space and a numerical factor that depends on the
definition
of the Chern-Simons term one obtains
\begin{equation} C_2=\int_{\Sigma}(A_1+\alpha')B^1=2\int_{\Sigma}(A_2 B^2
+A_3 B^3)(B^1+1)^{-1}B^1 \; , \end{equation}
the integrand of which coincides up to a power of $B^1$ with the integrand
of the observable Q. This is, however,
no coincidence but related to the fact that the Chern-Simons term is the
generating functional of the magnetic fields which play an important role
in the Ashtekar framework (see \cite{16}).\\

\section{Unitary transformation to the triad representation}

Now that we have the physical Hilbert space it is interesting to see
what it looks like in the triad representation because this
representation lies nearest to
the old metric formulation of quantum gravity and thus is more suitable
to an interpretation.
The most straightforward way to find the physical states in this
representation
turns out to solve simply the BRST-set of constraints for the connections
and apply the theorem of section (4.3.1).\\
Inspecting formulas (4.17)-(4.19) and recalling that on the constraint
surface
the magnetic fields are real we find that the constant c is real. In the
following, let us absorb
the sign ambiguity in taking the square root of c into $\sqrt{c}$.
Then, using 'cylinder coordinates' $(A_1,\sqrt{A},\alpha)$ and
$(E^1,\sqrt{E},\eta)$,
the 1-parameter-set of constraint functionals
\begin{eqnarray}
\zeta_1=E^1-\frac{c}{(B^1)^2},  \\
\zeta_2=E^2+\frac{2c B^2}{(B^1)^3}, \nonumber \\
\zeta_3=E^3+\frac{2c B^3}{(B^1)^3} \nonumber
\end{eqnarray}can be replaced by the constraint functionals at
most linear in the connection coefficients $A_I$
(4c has been replaced by c in this section)\begin{eqnarray}
\tilde{\zeta}_1 & = &
A_1+\eta'+(\arcsin(\frac{(E^1)'}{\sqrt{E(2+\sqrt{\frac{c}{E^1}})}}))'
\nonumber \\ & & +\frac{1}{4 c(2+\sqrt{\frac{c}{E^1}})}(\sqrt{\frac{c}
{E^1}})^3\sqrt{E(2+\sqrt{\frac{c}{E^1}})-((E^1)')^2},
\\\tilde{\zeta}_2 & = & A_2+\frac{E^3(E^1)'}{E}-\frac{E^2}{E}\sqrt{E(2
+\sqrt{\frac{c}{E^1}})-((E^1)')^2}, \\\tilde{\zeta}_3 & = &
A_3-\frac{E^2(E^1)'}{E}-\frac{E^3}{E}\sqrt{E(2+\sqrt{\frac{c}{E^1}})
-((E^1)')^2} \; .\end{eqnarray}The eqs. $\tilde{\zeta}_I=0$ can be
interpreted as the defining equations for the extrinsic curvature
or second fundamental form because, recalling formula (2.23), in each
equation the
2nd term on the rhs is just the negative of the corresponding component
of the spin-connection.
The theorem of section
(4.3.1) applies because $\tilde{\zeta}_I=0$ is a solution of a first class
set of constraints of the same number as there are configuration variables
(on the unconstrained phase space).\\It is well-known
that the spin connection is integrable (see \cite{2}; in fact, this property
is the reason for the Ashtekar-transformation to be a symplectomorphism) and
its generating functional reads
here\begin{equation} {\cal F}:=-\int_\Sigma dx\eta'E^1=\int_\Sigma dx\Gamma_1
E^1\; ; \; \eta=\arctan(\frac{E^3}{E^2}).  \end{equation}It is precisely the
reduction to spherical symmetry of that for
full gravity (compare \cite{2}), which can easily be checked by inserting the
expressions given in section 2.\\Recalling that in the triad representation
$A_I=-\delta/\delta E^I$and defining
$\Phi:=\ln(\tilde{\Psi})-\cal F$ we arrive at the following 'gradient
components' :\begin{eqnarray}\frac{\delta\Phi}{\delta E^1} & = &
(\arcsin(\frac{(E^1)'}{\sqrt{E(2+\sqrt{\frac{c}{E^1}})}}))'+\frac{\sqrt{
\frac{c}{(E^1)^3}}}{4(2+\sqrt{\frac{c}{E^1}})}\sqrt{E(2+\sqrt{\frac{c}
{E^1}})-((E^1)')^2}, \\\frac{\delta\Phi}{\delta E^2} & = &
-\frac{E^2}{E}\sqrt{E(2+\sqrt{\frac{c}{E^1}})-((E^1)')^2}, \\\frac{\delta
\Phi}{\delta E^3} & = & -\frac{E^3}{E}\sqrt{E(2+\sqrt{\frac{c}{E^1}})
-((E^1)')^2} \; ,\end{eqnarray}which can, by the method
presented in section (4.3.1), be integrated to give the 'potential'
\begin{eqnarray}& & \Phi[E^I]  =  \int_{\Sigma}dx E^I(x)\int_0^1 dt
\frac{\delta\Phi[F]}{\delta F^I}_{|F^J=t E^J}=
-\int_{\Sigma}dx\int_0^1 dt[(E^1)'\arcsin(\frac{(E^1)'}{\sqrt{E(2+
\sqrt{\frac{c}{t E^1}})}}) \nonumber \\ & & -\frac{1}{4(2\sqrt{t}+
\sqrt{\frac{c}{E^1}})}\sqrt{\frac{c}{E^1}}\sqrt{E(2+\sqrt{\frac{c}{t
E^1}})-((E^1)')^2}     +\sqrt{E(2+\sqrt{\frac{c}{t E^1}})-((E^1)')^2}] \; .
\end{eqnarray}Introducing the new variables \begin{equation} R:=\sqrt{E},\;
Y:=\frac{(E^1)'}{R},\;\rho^2:=2+\sqrt{\frac{c}{t
E^1}} \; , \end{equation}$\Phi$ takes the form\begin{equation}\Phi[E^I]=
-\int_{\Sigma}R dx\int_0^1 dt[Y\arcsin(\frac{Y}{\rho})+\sqrt{E(2+
\sqrt{\frac{c}{t E^1}})-Y^2}
-\frac{1}{4\sqrt{t}\rho}\sqrt{\frac{c}{E^1}}\sqrt{\rho^2-Y^2}] \; ,
\end{equation}in which the two first terms are easily recognized as the
integral of $\arcsin(Y/\rho)$ with respect to Y. On
the other hand (note that $\rho=\rho(t)$) we have\begin{equation} \frac{d}
{dt} \arcsin(\frac{Y}{\rho})=\frac{Y}{t\sqrt{\rho^2-Y^2}}\frac{1}{4\sqrt{t}
\rho}\sqrt{\frac{c}{E^1}}
 =\frac{d}{dY}(-\frac{1}{4\sqrt{t}\rho}\sqrt{\frac{c}{E^1}}
\sqrt{\rho^2-Y^2}) \; ,\end{equation}whence\begin{eqnarray}\Phi[E^I] & = &
 -\int_{\Sigma}R dx\int_0^1 dt\int_{Y_0}^Y
ds[\arcsin(\frac{s}{\rho})                 \frac{d}{ds}(-\frac{1}{4\sqrt{t}
\rho}\sqrt{\frac{c}{E^1}}\sqrt{\rho^2-s^2}] \nonumber \\ & = &
-\int_{\Sigma}R dx\int_0^1 dt\int_{Y_0}^Y
ds\frac{d}{dt}(t\arcsin(\frac{s}{\rho(t)})  = -\int_{R^1}R dx\int_{Y_0}^Y
ds\arcsin(\frac{s}{\rho(1)}) \nonumber \\ & = & -\int_{\Sigma}
dx[(E^1)'\arcsin(\frac{(E^1)'}{\sqrt{E(2+\sqrt{\frac{c}{E^1}})}})
 +\sqrt{E(2+\sqrt{\frac{c}{E^1}})-((E^1)')^2}] \; ,\end{eqnarray}where
the lower boundary  $Y_0$ was chosen appropriately.\\Hence
${\cal H}_{phys}$ in the triad representation is spanned by the following
 states :\begin{equation}\tilde{\Psi}_c[E^I]=\exp(-\int_{\Sigma}
dx[-E^1\Gamma_1+(E^1)'\arcsin(\frac{(E^1)'}{\sqrt{E(2+
\sqrt{\frac{c}{E^1}})}})
+\sqrt{E(2+\sqrt{\frac{c}{E^1}})-((E^1)')^2}] \}.\end{equation}
Thus, up to $\Gamma_1$, we are now
able to write the theory in metric language, because $q_{\theta\theta}=E^1,
E=2 E^1 q_{xx}=2 q_{\theta\theta}q_{xx}$. One may easily check
that the fall-off conditions derived in section 3 need further
specification in order to make the integrand of $\Psi_c[E^I]$ converge.
We refrain from giving it here because, at the moment, we are only
interested in formal manipulations for the triad
representation.\\Note that in the triad representation the constraint
functionals {\em cannot be chosen at most linear} with respect to the
momenta $A_I$ because the constraint
functionals of the canonical set are
{\em inhomogenous in $A_I$}. Nevertheless, the physical states can
be computed {\em if one uses the classical solutions of the constraints}
((6.2)-(6.4)) as the appropriate set of constraint
functionals.\\
It is in this sense that one can argue that a
1-parameter family of exact solutions of the Wheeler-Dewitt-equation has
been found for the spherically symmetric case. It can be verified that
these states are indeed annihilated by the Wheeler-DeWitt
constraint operator in its original form if one point-splits the second
functional derivatives and omits the factor $\exp(E^1\circ\Gamma_1)$.\\
The conclusion of all this might be that the usual
quantization procedure should be modified in such a way that one
{\em first} chooses a set of constraint functionals that contain the
momenta at most linearly and {\em then} applies the Dirac-procedure
thereby eliminating the operator-ordering difficulties right from the
 beginning.\\We will now sketch the proof that the above states can be
interpreted as formal Laplace-transforms of $\exp(icQ)$,
i.e.\begin{equation} \tilde{\Psi}_c[E^I]=\int[dA_1\wedge dA_2\wedge dA_3]
\exp(A_I\circ E^I)\exp(icQ). \end{equation}That this is only a formal
transformation corresponds to the fact that one has to
define non-analytical functions of operators by their spectral-resolution.
\\Solving $\zeta_I$ for $A_I$ to obtain $\tilde{\zeta}_I$ means that there
exists an (operator-valued) matrix $M_I\;^J$ such
that $M_I^J\tilde{\zeta}_J=\zeta_I$.
Let us define $\Psi_c[A_I]$ formally by\begin{equation}
\tilde{\Psi}_c[E^I]=\int[dA_1\wedge dA_2\wedge dA_3]\exp(A_I\circ E^I)
\Psi_c[A_I] \; . \end{equation}Then by the very definition of
$\tilde{\Psi}_c$ and a functional integration by parts($\lambda$ is
a suitable test function) we get\begin{eqnarray}0 & = &\lambda
M_I^J\circ\tilde{\zeta}_J\tilde{\Psi}_c[E^I]  \\  & = & \lambda\circ
\zeta_I(\hat{A}_J=-\frac{\delta}{\delta E^J},\hat{E}^J=E^J)
\tilde{\Psi}_c[E^I] \nonumber \\  & = & \int[dA_1\wedge dA_2\wedge dA_3]
\exp(A_I\circ
E^I)\lambda\circ\zeta_I(\hat{A}_J=A_J,E^J=\frac{\delta}{\delta A_J})
\Psi_c[A_I] \; . \nonumber\end{eqnarray}Inverting the Laplace-transform
shows that\begin{equation}
\lambda\circ\zeta_I(\hat{A}_J=A_J,E^J=\frac{\delta}{\delta A_J})
\Psi_c[A_I]=0 \end{equation}for any (suitable) test function $\lambda$.
But in section (4.3.1) we showed that these 3 equations have the
(up to a constant factor) solution $ \Psi_c(A_I)= \exp(-icQ)$,
which completes the proof. Unfortunately we were not able to give a
direct proof by doing the functional integral (6.15)
explicitly. However, one can show that a saddle-point approximation
gives the correct result.\\The scalar product for the triad representation
can be taken over from the self-dual representation simply
by defining a measure that makes the Laplace-transformation unitary.
This measure $\mu$ turns out to be non-local as was to be expected
recalling the general results of ref. \cite{19} because the
Ashtekar-transformation is a {\em complex }  symplectomorphism (see
section (4.4) for notation) :\begin{eqnarray} <\Psi_1|\Psi_2 > & := &
\int[d(E_1)^I][d(E_2)^I]\mu[(E_1)^I,(E_2)^I]\bar{\tilde{\Psi}}_1
[(E_1)^I]\tilde{\Psi}_2 [(E_2)^I]\;,    \\ \mu[(E_1)^I,(E_2)^I] & := &
\int[dE^J dA_J dc^J d\rho_J]\exp(\{G,\Omega\}+A_J\circ
(E_2)^J+\overline{A_J\circ (E_1)^J}) \; . \nonumber\end{eqnarray}
We have the usual interpretation (due to positive definiteness of the
scalar product) that for any $f\in L_2(R,dc)$ and physical
state\begin{equation} \Psi=\int_R dc f(c)\Psi_c \end{equation}
$\tilde{\Psi}[E^I]$ is the probability amplitude for the pure
spherically symmetric gravitating sytem to adopt the field configuration
$E^I$ in the state $\Psi$. The restriction on f ensures that $\Psi$ is
 normalizable :\begin{equation}<\Psi|\Psi>=\int_R dc_1\int_R dc_2
<\Psi_{c_1}|\Psi_{c_2}>\bar{f}(c_1)f(c_2)=2\pi\int_R dc
|f(c)|^2\end{equation}because \[ <\Psi_{c_1}|\Psi_{c_2}>=\int_R dQ
\exp(i(c_1-c_2)Q)=2\pi\delta(c_1-c_2) \; . \]As an application let
us compute the extrema of $|\tilde{\Psi}[E^I]|^2$ for a given state
$\Psi$. For a general f this is not easy. However, we are interested
mainly in the eigenstates of observables. Now the observable P has the
simple eigenstates $\Psi_c(Q)=\exp(icQ)$ and the spectrum is
the real line : $c\in R$, because P is self-adjoint. Since $\Psi_c$ is
not normalizable we construct the state $\hat{\Psi}_c:=\int_R dc' f_c(c')
\Psi_{c'}$, where $f_c$ may be, for instance, a Gaussian
function as sharply peaked around c as we like. We then compute the
extrema of $|\tilde{\Psi}_c[E^I]|^2$, which approximate the extrema of
the probability density of $\hat{\Psi}_c$ as well as we
like. Before doing this one has to take into account the following fact :
The pre-factor $\exp(\int_\Sigma dx E^1\Gamma_1)$ common to {\em all}
physical states must not be varied when determining the
extrema of the absolute square of the states because it is the result of
imposing the reality condition $A_I+\bar{A}_I=2\Gamma_1$ on the states and
thus eventually belongs to the measure of the triad
scalar product which could be defined as
follows\begin{equation} <\Psi|\Phi>:=\int_R[dE^1\wedge
dE^2\wedge dE^3]\bar{\Psi}[E^J]\exp(2\int_\Sigma dx
\Gamma_1 E^1)\Phi[E^J] \; , \end{equation}and implies
that $(A^\dagger)_I+A_I=2\Gamma_I$. The latter result is again in nice
agreement with ref. \cite{19}.\\The solution of our variational
problem is then\begin{equation}
\sqrt{2+\sqrt{\frac{c}{E^1}}}=\pm\frac{(E^1)'}{\sqrt{E}} \; .\end{equation}
The classical Schwarzschild solution is contained in this class of field
configurations : choose $\displaystyle E^1=r^2,E=2
E^1(1-\frac{2m}{r})^{-1},c=16 m^2$ and the negative sign for $\sqrt{c}$.
\\Note that the integrand in the exponential of eqn. (6.14) reduces
to $-E^1\Gamma_1$ if eqn. (6.23)
holds.\section{Conclusions}Ashtekar's quantization programme for gravity
could be carried out completely in the self-dual representation with
restriction to spherical symmetry. Although the model has
many aspects common with full gravity, there is one property that is
not shared by full gravity and which simplified important steps of the
quantization procedure used by us, namely that the constraint functionals
could be replaced by constraint functionals at most linear in the momenta.
It should also be mentioned that in such a situation the reduced phase space
method and the operator constraint method give equivalent results.\\
There may be other elements of our analysis
which could be applied to more general systems, e.g. the treatment of
observables or the construction of the scalar product.\\ It also
turned out that the operator ordering is of physical significance and
that care must be taken when regularizing full quantum gravity in order
not to destroy physically important properties of the theory
by point-splitting. Even more interesting, the operator ordering implied
that the classical structure functions do not resemble their quantum
counterparts. This latter result should be valid for the
quantization of
full gravity in the self-dual representation, too.\\Finally,
it is remarkable that not the Ashtekar-constraints, but the BRST-constraints
 seem to be the natural ones. This might indicate
that a rearrangement of Ashtekar's constraints is useful in other models,
too. Moreover, as explained in section 6, it could well be that the
quantization in a general representation via the
Dirac-procedure is possible only, if one first computes the constraint
surface explicitely, i.e. casts the constraints into a form in
which the momenta appear at most linearly if one does not want to make
use of point-splitting.\\ \\{\large Acknowledgements} \\ \\We thank
R. Loll for a seminar which stimulated our interest in Ashtekar's
formulation of general relativity and we are indebted to
A.\ Ashtekar for helpful discussions during the spring meeting
of the German Physical Society in Berlin, where the main results
of this paper were presented. Finally we thank the referee for suggesting
several  improvements of the first  version of the manuscript.

\begin{appendix}

\section{Appendix}

In the present paper, 2 sets of constraint functionals are used. They are
given
here only for the more physical sector I. Also
they are written down directly in the correct ordering for quantum theory.
In order to obtain the corresponding Poisson algebras one simply has to turn
off the ordering and to multiply the rhs by a factor of (-i). For
completeness,
we also give the fall-off behaviour of the corresponding
Lagrange-multipliers.\\
All appearing integrals have $\Sigma$ as domain of integration.
The various surface terms all vanish, as classical functions as well as
operators
on ${\cal H}_{phys}$, on which they are null-operators, due to the fall-off
behaviour of the test-functions defined in section 3. This can be seen as
follows : \\
In the following calculations the first and last line contain operators $O_1$
and $O_2$
that annihilate ${\cal H}_{phys}$. Suppose that during the calculation we
neglected
a surface operator S, i.e. $O_1=O_2+S$. Then $S{\cal H}_{phys}=0$. Since S
is a
surface term, it cannot be proportional to the constraint functionals, so it
must vanish
identically on ${\cal H}_{phys}$ and thus may be neglected. The technical
reason is that the fields (together with the multipliers) that appear in
the surface term after applying S to physical states fall off strongly enough
at infinity, so that the surface term vanishes.\\
\\
i) Canonical set
\begin{eqnarray}
{\cal G} & = & (E^1)'+A_2 E^3-A_3 E^2, \\
H_x  & = & -A_1 (E^1)'+A_2'E^2+A_3'E^3, \nonumber \\
H & = & \frac{1}{2D}(E^2 D [2 B^2 E^1+B^1 E^2] + E^3 D [2 B^3 E^1+B^1 E^3]),
\nonumber \\
(\Lambda,N^x,\tiN) & = & (O(r^{-2-\epsilon}),O(r^{-\epsilon}),O(r^{-2
-\epsilon})) \; ,\nonumber\end{eqnarray}
where D was defined in connection with eqs. (4.37).

ii) BRST-set
\begin{eqnarray}
\phi_1 & = & 2 ((B^1)' E^1+B^1(E^1)'),  \\
\phi_2 & = & E_2+2\frac{B^2}{B^1}E^1, \nonumber \\
\phi_3 & = & E_3+2\frac{B^3}{B^1}E^1, \nonumber \\
(\lambda^1,\lambda^2,\lambda^3) & = & (O(r^{-1-\epsilon}),O(r^{-2
-\epsilon}),O(r^{-2-\epsilon})) \; , \nonumber
\end{eqnarray}
where the mutipliers $\lambda^I$ are defined to be the coefficients of the
constraint functionals of the BRST-set when replacing the canonical set by
the $\phi_I$ in the action. The algebra is again of
rank 1 and the BRST-charge computed by standard methods (see \cite{8}) is
given by
($c^I,\rho_I$ are ghosts and anti-ghosts)
\begin{eqnarray}
\Omega & = & c^A\circ\phi_A+\frac{1}{2}U[c^A,c^B]_{AB}\;^C\circ\rho_C
\nonumber \\
 & = & c^A\circ\phi_A-\frac{i}{2(B^1)^2}(2 B^1 c^1(A_2 c^2+A_3 c^3)+2
 c^2 c^3)\circ\rho_1 \; .
\end{eqnarray}
\\ \\

The algebra of the BRST-set is as follows:\\
\\
\begin{equation} [M\circ\phi_1,N\circ\phi_1]=0 \; , \end{equation}
since only the momentum $E^1$ but no $A_1$ is contained in $\phi_1$.
Furthermore
\begin{eqnarray}
 & & [M\circ\phi_2,N\circ\phi_2]  \\
 & = & \int dx\int dy (M(x)N(y)-M(y)N(x))([E^2(x),\frac{2 B^2}{B^1}(y)]
 E^1(y) \nonumber \\
 &   & +\frac{2 B^2}{B^1}(x)[E^1(x),\frac{2 B^2}{B^1}(y)]E^1(y))=0 \nonumber
\end{eqnarray}
since no spatial derivative survives in either of the commutators.
Similarily we have
\begin{equation} [M\circ\phi_3,N\circ\phi_3]=0 \; . \end{equation}
The calculation of the rest of the commutators is slightly more difficult:
\begin{eqnarray}
 & & [M\circ\phi_1,N\circ\phi_2] \\
 & = & \int dx\int dy M(x)N(y)(2[A'(x),E^2(y)]E^1(x)+[(A-2)(x),E^2(y)]
 (E^1)'(x)\nonumber \\
 &   & +(2A'(x)[E^1(x),B^2(y)]+(A-2)(x)[(E^1)'(x),B^2(y)])\frac{4}{A-2}
 (y)E^1(y)) \nonumber \\
 & = & \int dx(4N A_2(M E^1)'-2 M N A_2 (E^1)'+(M N 2A'A_2-N(M(A-2))'A_2)
 \frac{4}{A-2}E^1) \nonumber \\
% & = & \int dx(4N A_2(M E^1)'-2 M N A_2 (E^1)'+M N A'A_2\frac{4}{A-2}
E^1-4 N M'A_2 E^1) \nonumber \\
% & = & \int dx(2 M N A_2 (E^1)'+M N A'A_2\frac{4}{A-2}E^1)
%  =  \int dx M N \frac{2 A_2}{A-2}((A-2)(E^1)'+2A'E^1) \nonumber \\
% & = & M N \frac{2 A_2}{A-2}\circ\phi_1
 & = &  M N \frac{A_2}{B^1}\circ\phi_1. \nonumber
\end{eqnarray}
Similarily, or by O(2)-symmetry :
\begin{equation} [M\circ\phi_1,N\circ\phi_3]=M N \frac{A_3}{B^1}\circ\phi_1
\; .
\end{equation} Finally
\begin{eqnarray}
 & & [M\circ\phi_2,N\circ\phi_3] \\
 & = & \int dx\int dy M(x)N(y)([E^2(x),\frac{2 B^3}{B^1}(y)]E^1(y)-[E^3(y),
 \frac{2 B^2}{B^1}(x)]E^1(x) \nonumber \\
 &   & +\frac{2 B^2}{B^1}(x)[E^1(x),B^3(y)]\frac{2}{B^1}(y)E^1(y)-\frac{2
 B^3}{B^1}(y)[E^1(y),B^2(x)]\frac{2}{B^1}(x)E^1(x)) \nonumber \\
 & = & \int dx(-8 M N\frac{B^3}{(A-2)^2}A_2 E^1+8 M N\frac{B^2}{(A-2)^2}
 A_3 E^1+4 M(N\frac{1}{A-2}E^1)' \nonumber \\
 &  & +4 N(M\frac{1}{A-2}E^1)'+16 M N\frac{B^2}{(A-2)^2}A_3 E^1-16 M N
 \frac{B^3}{(A-2)^2}A_2 E^1) \nonumber \\
 & = & \frac{M N}{(B^1)^2}\circ\phi_1 \; . \nonumber
\end{eqnarray}
\\ \\

As to the algebra of the canonical set
it is impossible to choose the above mentioned function D in such a way
that the structure 'functions' for the canonical algebra become the
classical ones
given in eqs. (2.18) although, of course, these are recovered
in the classical limit. Moreover, the structure functions become
{\em nonanalytical in the momenta}. This is, of course, quite undesireable.
However, it is the price to pay in order that the constraint operators always
appear to the right, i.e. that the algebra strictly closes. In this way
one never has to make
sense of a functional derivative in the denominator. Nevertheless it is
obvious that the canonical set is not the natural one for the model.\\
The commutators of the generators of the kinematical subgroup $O(2)\times Diff
\Sigma$
yield the same structure functions as their classical counterparts, because
they are at most linear in the momenta. The remaining commutators, however,
are rather
involved for a general D and we refrain from displaying them here, but rather
wish to give a rigorous general argument showing the closure of the
algebra in the operator ordering defined.\\
The following list of equations gives the 'transformation matrix' for sector
I between the 2 equivalent
sets of constraints, where the sequence of the operators is important:
\begin{eqnarray}
\chi_1 & := & {\cal G}=\frac{1}{B^1}\phi_1+A_2\phi_3-A_3\phi_2, \\
\chi_2 & := & H_x=B^2\phi_3-B^3\phi_2, \nonumber \\
\chi_3 & := & H=\frac{1}{2 D}(E^2 B^1 D\phi_2+E^3 B^1 D\phi_3) \nonumber \\
 & \iff & \nonumber \\
\phi_1 & = & B^1 {\cal G}-A_2\phi_3+A_3\phi_2\; \mbox{, where}  \\
\phi_2 & = & \frac{1}{E^2 B^2+B^3 E^3}\frac{B^2}{B^1 D}[2 D H-E^3
\frac{B^1 D}{B^2}H_x], \nonumber \\
\phi_3 & = & \frac{1}{E^2 B^2+B^3 E^3}\frac{B^3}{B^1 D}[2 D H+E^2
\frac{B^1 D}{B^3}H_x]\; . \nonumber
\end{eqnarray}
The idea is now as follows : let the operator-valued matrices $c_I^J$ and
$d_I^J$
be defined by $\chi_I=:c_I^J\phi_J$ and $\phi_I=:d_I^J\chi_J$ and let the
structure
'functions' of the BRST set as derived above be denoted by U. Then
\begin{eqnarray}
 & & [M\circ\chi_I,N\circ\chi_J] \nonumber \\
 & = & [M\circ c_I^K\phi_K,N\circ c_J^L\phi_L] \nonumber \\
 & = & M c_I^K\circ[\phi_K,N\circ c_J^L\phi_L]+[M c_I^K,N\circ\chi_J]\circ
 \phi_K \nonumber \\
 & = & M c_I^K\circ[\phi_K,N c_J^L]\circ\phi_L+M c_I^K\circ N c_J^L\circ[
 \phi_K,\circ\phi_L]
 +[M c_I^K,N\circ\chi_J]\circ\phi_K \nonumber \\
 & = & (M c_I^K\circ[\phi_K,N c_J^P]+U[M c_I^K,N c_J^L]_{KL}\;^P+[M c_I^P,
 N\circ\chi_J])\circ\phi_P \nonumber \\
 & = & (M c_I^K\circ[\phi_K,N c_J^P]+U[M c_I^K,N c_J^L]_{KL}\;^P+[M c_I^P,
 N\circ\chi_J])d_P^Q\circ\chi_Q \; .
\end{eqnarray}
Finally, in order to see what problems arise with the canonical set we choose
D=1 and compute the commutator between 2 scalar constraints. The result is
\begin{equation} [M\circ H,N\circ H]=\frac{1}{2}(M N'-M'N)\circ E^1(E^3 B^1
\phi_2-E^2 B^1\phi_3) \; , \end{equation}
where the two constraints of the BRST-set on the rhs must be expressed, via
the above
transformation, in terms of the canonical set. Clearly, the structure
'functions'
are then not polynomial, not even analytic, in the momenta although the
constraints
stand always on the right. The classical limit of the operator on the rhs
of (A.13)
is the expected one: $(E^1)^2 H_x$ (see eqs. (2.18)).\\
\\
\\
We finally show the weak commutativity of P and Q with the constraint
functionals
of the canonical set.\\
As the BRST-set of constraints is at most linear in the momenta, the
annihilation of Q
by the (smeared) BRST-constraints therefore implies that it strongly commutes
with these constraints. So far for Q. For P we get
\begin{equation} [P,M\circ\phi_1]=0 \end{equation}
trivially, since no $A_1$ but only $E^1$ is contained in both operators.
Furthermore
\begin{eqnarray}
 & & [P,M\circ\phi_2]  \\
 & = & \int_\Sigma dx\int_\Sigma dy M(y) (B^1(x)^2[E^1(x),B^2(y)]\frac{2}
 {B^1(y)}E^1(y)+[B^1(x)^2,E^2(y)]E^1(x)) \nonumber \\
 & = & \int_\Sigma dx M (2 B^1 A_2 E^1-2A_2 B^1 E^1)=0  \; , \nonumber
\end{eqnarray}
and similarily
\begin{equation} [P,M\circ\phi_3]=0 \; . \end{equation}
So the BRST-constraint functionals even {\em strongly} commute with both
observables.
Proving weak commutativity with all the canonical constraint functionals
directly, only
using the CCR, is hard work. We will give a more elegant indirect proof,
based
on the following argument which exploits the existence of the transformation
(A.10) and (A.11)
between the two sets of constraint functionals. Adopting the same notation as
above we have for any observable O with respect to the BRST-set
\begin{eqnarray}
[O,N\circ\chi_I] & = & N c_I^J\circ [O,\phi_J]+[O,N c_I^J]\circ\phi_J  \\
& = & (N c_I^J V[O]_J^K+[O,N c_I^K])\circ\phi_K \nonumber \\
& = & (N c_I^J V[O]_J^K+[O,N c_I^K])d_K^L\circ\chi_L \; , \nonumber
\end{eqnarray}
where the structure functions of O are defined by
\begin{equation} [O,M\circ\phi_I]:=V[O]_I^J M\circ\phi_J:=\int_\Sigma dx
\int_\Sigma dy V[O]_I^J(x;y)M(x)\phi_J(y) \; , \end{equation}
and vanish for our observables P and Q. This completes the proof.\\
For completeness sake we will display the {\em Poisson-bracket} of the
observable Q with the scalar constraint :
\begin{eqnarray}
& & \{M\circ{\cal H},Q\}=\int dx M[\frac{2}{A'}f((B^2 A_3'-B^3 A_2')
{\cal G}-(B^2 A_2+B^3 A_3)\xi)) \nonumber \\
 &  & +\frac{\dot{f}}{(E^1)'}({\cal H}-\frac{A-2}{(A')^2}((A_3'{\cal G}
 -A_2\xi)^2+(A_2'{\cal G}+A_3\xi)^2) \nonumber \\
 &  & +\frac{2}{(A')^2}(A' E^1+(A-2)(E^1)')((B^2 A_3'-B^3 A_2'){\cal G}
 -(B^2 A_2+B^3 A_3)\xi)] \; ,
\end{eqnarray}
which shows that one never would have found Q by using the original
definition of
an observable, that is $\{Q,M\circ\chi_I\}\approx 0$. Furthermore it
is obvious
that the canonical set is completely inappropriate for the construction
of the
scalar product as presented in section (4.4), because with these complicated
first order structure functions its BRST-extension is probably 'long'.

\end{appendix}

\end{document}